\documentclass[11pt]{article}
\usepackage{mathtext}
\usepackage[T2A]{fontenc}
\usepackage[cp1251]{inputenc}
\usepackage{amssymb}
\usepackage{graphics}
\usepackage[dvips]{graphicx}
\usepackage{color}
\usepackage{amsbsy}
\usepackage{citehack}
\usepackage{longtable}
\usepackage[left=2cm,right=2cm,top=2cm,bottom=2cm,bindingoffset=0cm]{geometry}
\usepackage{amsmath}
\usepackage{ccaption}
\usepackage{epstopdf}
\usepackage{soul}
\usepackage{verbatim}
\usepackage{dsfont}

\title{Dynamic fracture of a discrete dissimilar chain:\\ transient, subsonic and supersonic regimes.}

\author{Nikolai Gorbushin, Gennady Mishuris
\\
{\it Department of Mathematics,
Aberystwyth University, }
\\ {\it Ceredigion SY23 3BZ, Wales, UK}
}
\date{}

\begin{document}

\maketitle

\begin{abstract}
This paper deals with the theoretical and numerical analysis of dynamic fracture of dissimilar chain consisting of masses lined by springs. Such a structure exhibits quite different dynamic properties in comparison with a symmetrical uniform structure when dynamic properties are in question. Among other to stay in a balance, the external force applied to the system should not be the same but depend on the seed of the crack propagating
as the result of the force action. Moreover, in the supersonic regime there is a bang gap in the velocity where the crack of that speed cannot propagate at all. However, having such theoretical prediction, a question still remains where and how those dynamical regimes can be achieved in real structure. We answering on this question by providing tailored numerical simulations demonstrating that various predicted steady-state regimes can be reached after rather short transient states. Among other, we analise how position of applied loading influence the result.
\end{abstract}


\section{Introduction}

The investigation of particularities in dynamic fracture of bi-materials is approached from the experimental, theoretical and computational points of view. One of the most addressed questions in dynamic problems of fracture is to establish the limits of the crack propagation. The limiting crack speed in homogeneous solids is predicted to be a fraction of Rayleigh speed~\cite{yoffe1951}, and series of experiments~\cite{dally1979,livne2008,ravi1984} and numerical simulations~\cite{abraham2000,xu1994} also demonstrate the growth of instabilities while crack starts to move at high speed. Although the variety of scientific works on dynamic crack propagation in homogeneous solids is high, papers on the interfacial cracks in bi-material is not are not that numerous. The last are able to discover some new phenomena and also provide the possible tests of existed fracture theories. Some of the theoretical and experimental works in this field should be mentioned.

The experimental data achieved by means of a weight tower device and a gas gun for the dynamic crack growth is presented in~\cite{lambros1995}. The authors also attempted to develop a fracture condition based on the complex stress intensity factors for the crack speeds within a subsonic range. The experimental observations are supported by the high-order asymptotic analysis~\cite{liu1993}. The asymptotic analysis of interfacial steady-state crack growth in~\cite{xiaomin1994} proposes to use stress intensity factor rather than an energy release rate for these problems as the last vanishes and may cause the computational unrealistic predictions. The contact between the solids behind a steadily growing crack along the interface in a dissimilar solid is taken into an account in~\cite{huang1998}. It made possible to explain the experimental observations, shown in the same paper, more confidently. High contrast in material properties of a dissimilar solid unveil the intersonic crack speeds~\cite{rosakis1998} which are predicted to be forbidden in homogeneous solids.

The closer look at the interface between the solid can make new discoveries on fracture process of bi-materials. For that, the discrete representation of the solid is required. The theoretical works on this topic are limited due to the complexity of involved mathematical techniques and mostly addressed through numerical simulations~\cite{buehler2004,buyukozturk2011}. The theoretical study~\cite{tewary1992} concerns the bi-material solids in the framework of lattice structures but it only considers a quasi-static formulation. The steady-state fault propagation in a dissimilar structure is analysed in~\cite{mishuris2012}, where the technique developed by Slepyan for the crack propagation in lattice structures has been utilised~\cite{slepyan1984,slepyan1981,slepyan2012}. The method to solve dynamic fracture problems in discrete structures proposed by Slepyan appears to be very efficient and applicable to many configurations. For instance, the cracks in triangular~\cite{marder1995,pechenik2002,slepyan2001} and square-cell lattices~\cite{mishuris2009,slepyan2001feeding}, one-dimensional models of prestressed structure~\cite{ayzenberg2014}, chain with non-local interactions~\cite{gorbushin2017analysis} and beam structure~\cite{nieves2016}.

In this work we study a one-dimensional model of a crack propagation in a double chain of two different materials linked together by elastic springs under mode III fracture. The crack occupies a semi-infinite region and is driven by the moving forces remotely located behind a steadily moving crack tip. The similar problem for the identical materials of two chains is thoroughly studied in~\cite{gorbushin2017dynamic}. Two different problems appear to be formally mathematically equivalent to that discussed in this paper: Mode II loading of a bi-material structural interface with buckling roads and fracture of dissimilar chain in Mode II configuration in~\cite{berinskii2017}.

\section{Problem formulation.}
\begin{figure}[!ht]
\center{\includegraphics[scale=0.8]{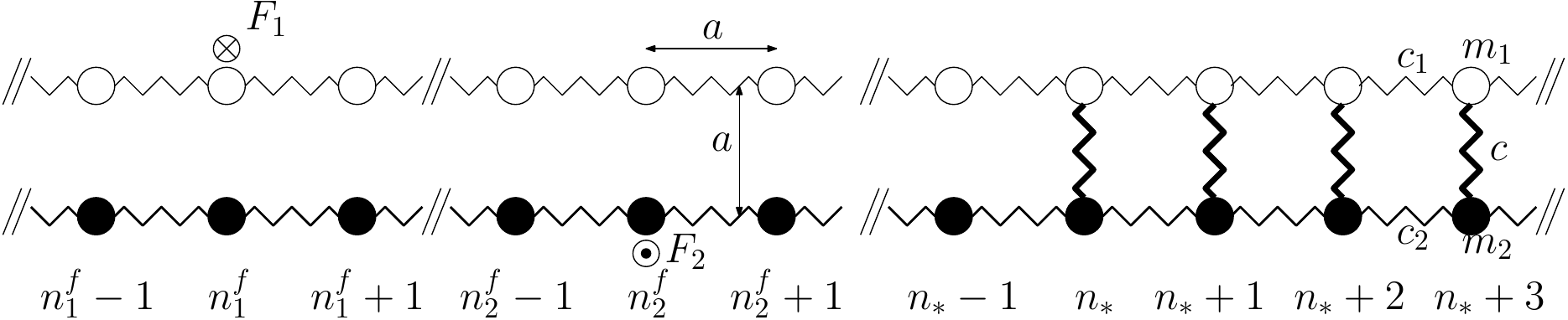}}
\caption[ ]{\small Infinite double chain consisting of dissimilar oscillators linked by massless springs. The stiffness of the springs ($c_1,c_2$) as well as the masses of the oscillators ($m_1,m_2$) are, generally speaking, different. Two chains are interconnected by linear springs with stiffness $c$ where the intact part of the structure starting from oscillators with index $n_*$ which represents "a crack tip" moving on the right with some speed $v$.}
\label{fig:DissimilarChain}
\end{figure}

Let us consider a structure of a double chain of dissimilar oscillators shown on Fig.\ref{fig:DissimilarChain}. The parameters of the model are: $m_1$, $m_2$ - masses of upper and lower chains, respectively, $c_1$, $c_2$ spring constants between the oscillators of the same sort. The oscillators are numbered with indices $n$. The masses with the same index $n\geq n_*$ are connected with each other by the linear springs
of the stiffness $c$. There are external forces $F_1$, $F_2$ applied to the upper and lower chains at points $n=n_1^f$ and $n=n_2^f$, respectively.

System of Newton equations representing the structure movement can be written in the form:
\begin{equation}
\begin{gathered}
\begin{gathered}
m_1\frac{d^2}{dt^2}u_n(t)=c_1(u_{n+1}(t)+u_{n-1}(t)-2u_{n}(t))+F_1\delta_{nn_1^f},\\
m_2\frac{d^2}{dt^2}w_n(t)=c_2(w_{n+1}(t)+w_{n-1}(t)-2w_{n}(t))-F_2\delta_{nn_2^f},
\end{gathered}\quad n<n^*,
\\
\begin{gathered}
m_1\frac{d^2}{dt^2}u_n(t)=c_1(u_{n+1}(t)+u_{n-1}(t)-2u_{n}(t))+c(w_n(t)-u_n(t)),\\
m_2\frac{d^2}{dt^2}w_n(t)=c_2(w_{n+1}(t)+w_{n-1}(t)-2w_{n}(t))+c(u_n(t)-w_n(t)),
\end{gathered}\quad n\geq n_*,\\
\end{gathered}
\label{eq:OriginalProblem}
\end{equation}

The oscillators with index $n_*=n_*(t)$ represent a crack tip. The movement of the crack is a result of the breakage of the respective link $n_*=n_*(t)$ at the corresponding moment $t=t_*$ such that the following conditions are valid:
\begin{equation}
|u_{n_*}-w_{n_*}|=\epsilon_c,\quad |u_{n}(t_*)-w_n(t_*)| < \epsilon_c, \quad n>n_*,
\label{fracture_cond_0}
\end{equation}
where $\epsilon_c=const$ is a strength quantity of the springs of stiffness $c$. Both chains separately have characteristic speeds of sounds that limit the signal propagation within the chains. Their values, normalised by equilibrium distance $a$, are:
\begin{equation}
v_{j}=\sqrt{\frac{c_j}{m_j}},\quad j=1,2.
\label{eq:SpeedsOfSound}
\end{equation}
For a sake of convenience, we introduce the following auxiliary values:
\begin{equation}
\beta_{j}=\sqrt{\frac{c}{m_{j}}},\quad j=1,2,
\label{eq:Beta_12}
\end{equation}
having the same dimension as speeds, $v_j$. We also assume that the applied forces move with constant speeds, $v_j^f$, thus, their positions in the structure change according to the following rule:
\begin{equation}
n_{j}^f(t)=v_{j}^ft,\quad j=1,2,
\label{eq:ForceSpeeds}
\end{equation}
which should also be rounded to the integer number.
% and $v_{j}^f,j=1,2$ are the velocities of top and bottom forces, respectively.

As a result of such action, the "crack tip" can propagate from the left to the right which is a discrete movement. Times between the consequent breakages may be different or equal. In the last case we say the crack
moves with a contact speed, $v$. It is well known~\cite{slepyan1984} that the moving crack tip generates waves which may propagate in both directions (on the right and on the left) away from the crack tip
and, when considering destroyed part of the structure,
the waves propagate with generally speaking different speeds, $v_1$ and $v_2$ along the top and the bottom part of the structure. Finally, we always assume that no incident waves brings energy into the structure from infinity.

The formulation should be supplemented by initial conditions describing the displacements and velocities of each consequent mass along the chain.
It is easy to observe that possible solution of the problem, $u_n(t)$, $w_n(t)$  can be found with accuracy to linear functions in the following manner:
\begin{equation}
u_n(t)=a_1t+a_0+n(b_1t+b_0)+u_n^\star(t),\quad w_n(t)=a_1t+a_0+n(b_1t+b_0)+w_n^\star(t).
\label{eq:Rigid_body}
\end{equation}
where $u_n^\star(t)$ and $w_n^\star(t)$ are the solution satisfying zero initial conditions.

Since the aim of the paper is to analyse possible steady state regimes we always apply zero initial conditions in the numerical simulations. However, since we also interested in the transient stages,
various initial conditions (as well as the initial position of the applied forces) may affect this part of the analysis. We will comment on this in the proper place.

\section{Quasi-static problem.}
\begin{figure}[!ht]
\center{\includegraphics[scale=0.8]{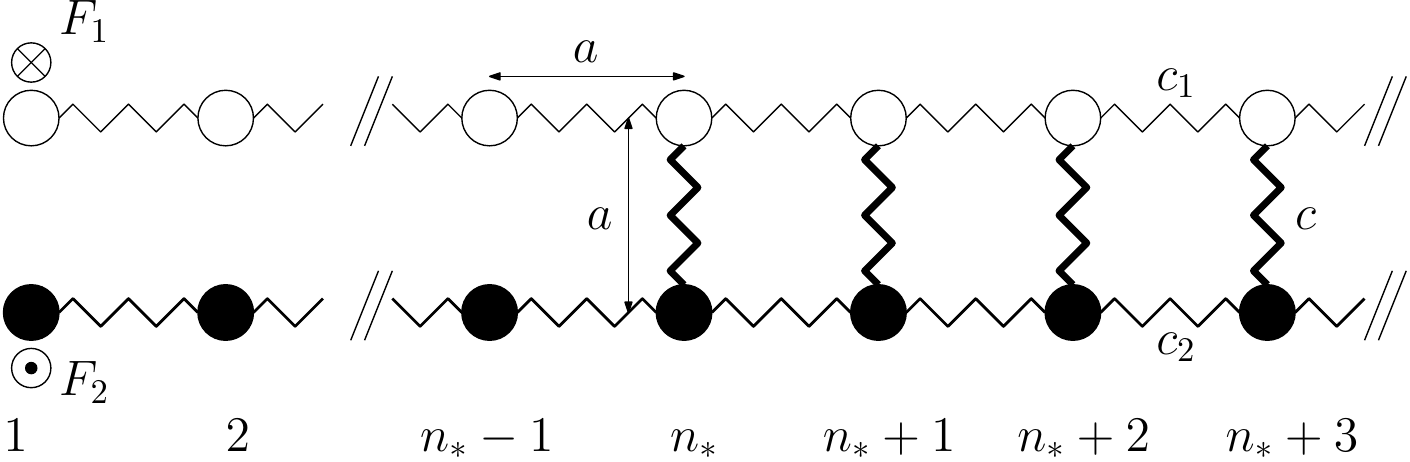}}
\caption[ ]{Double chain of dissimilar oscillators with $c_1,c_2$ being spring constants, respectively. Two chains are connected together by linear springs with stiffness $c$ starting from oscillators with index $n_*$ which represents a crack tip.}
\label{fig:StaticDissimilarChain}
\end{figure}

We start from the quasi-static analogue of the problem to derive some limiting relations for a stationary crack. We refer to the configuration of a faulted double chain displayed in Fig.\ref{fig:StaticDissimilarChain} that differs to the Fig. 1 only with the position of the loads $F_1$ and $F_2$ which are applied now at the same point $n=n_f$
which is simultaneously the left-hand end of the chain.
It is trivially to understand that, in the static case, the position of the load plays no role for an evident reason.
It is also straightforward conclusion that the external forces should be equal $F_1=F_2$ to guarantee the structure is in an equilibrium condition.
We will prove this later by introducing the conditions require the equilibrium:
\begin{equation}
u_n,w_n\to 0, \quad n\to+\infty.
\label{eq:Static_infinity}
\end{equation}

It easy to find the static solution by splitting the analysis to the separate study of the broken and intact parts of the structure.
For the latter one has:
\begin{equation}
\begin{gathered}
c_1(u_{n+1}+u_{n-1}-2u_{n})-c(u_{n}-w_{n})=0,\\
c_2(w_{n+1}+w_{n-1}-2w_{n})+c(u_{n}-w_{n})=0.
\end{gathered}
\quad n\geq n_*,
\label{eq:StaticChainIntact}
\end{equation}
It is convenient to present the solution in terms of the linear combinations, $\psi_n$ and $\phi_n$, of the displacements $u_n$ and $w_n$:
\begin{equation}
\psi_n=u_n-w_n,\quad \phi_n=u_n+w_n.
\label{eq:StaticPsiPhi}
\end{equation}
Notice, that function $\psi_n$ describe the force in the springs of stiffness $c$ between the corresponding masses for $n\geq n_*$ and crack opening for $n<n$, while the function
$\psi_n$ shows the deviation of the middle line of the structure from the symmetry line.

As a result, equations \eqref{eq:StaticChainIntact} can be then presented in the form:
\begin{equation}
\begin{gathered}
\psi_{n+1}+\psi_{n-1}-(2+\alpha)\psi_n=0,\\
\phi_{n+1}+\phi_{n-1}-2\phi_n=\beta\psi_n.
\end{gathered}
\quad\quad  n\geq n_*.
\label{eq:StaticChainIntact_psi_phi}
\end{equation}
Here we have introduced dimensionless parameters:
\begin{equation}
\alpha=\mu_1+\mu_2,\quad \beta=\mu_1-\mu_2, \quad \mu_j=\frac{c}{c_j}.
\label{eq:StaticContrasts}
\end{equation}

At the crack tip, $n=n_*$, the fracture condition \eqref{fracture_cond_0}$_1$ should be also imposed:
\begin{equation}
\psi_{n_*}=\epsilon_c,
\end{equation}
while at infinity we have \eqref{eq:Static_infinity}:
\begin{equation}
\psi_n,\phi_n\to 0, \quad n\to+\infty.
\label{eq:Static_infinity_1}
\end{equation}
The solution of equations \eqref{eq:StaticChainIntact_psi_phi} can be written in the form:
\begin{equation}
\psi_n=\epsilon_c\lambda^{n-n_*},\quad \phi_n=\epsilon_c\frac{\beta}{\alpha}\lambda^{n-n_*},\quad n\geq n_*,
\label{eq:StaticSolutionIntact}
\end{equation}
with factor $\lambda$ satisfying the condition $|\lambda|<1$, that should be found from the respective quadratic equation to be:
\begin{equation}
\lambda=\frac{\sqrt{4+\alpha}-\sqrt{\alpha}}{\sqrt{4+\alpha}+\sqrt{\alpha}}.
\label{eq:StaticLambda}
\end{equation}

Solution in the broken part of structure satisfies the system of equations:
\begin{equation}
\begin{gathered}
c_1(u_2-u_1)+F_1=0,\quad c_2(w_2-w_1)-F_2=0,\quad n=1,\\
\begin{gathered}
c_1(u_{n+1}+u_{n-1}-2u_{n})=0,\\
c_1(w_{n+1}+w_{n-1}-2w_{n})=0.
\end{gathered}
\quad 1<n<n_*,
\end{gathered}
\end{equation}
and can be found directly:
\begin{equation*}
u_n=\frac{F_1}{c_1}\left(n_*-n\right)+u_0,\quad
w_n=-\frac{F_2}{c_2}\left(n_*-n\right)+w_0,\quad 1\leq n<n_*,
\end{equation*}
where $u_0,w_0$ are the values of displacements at a crack tip. From this we conclude that:
\begin{equation}
\begin{gathered}
\psi_n=\left[\frac{F_1}{c_1}+\frac{F_2}{c_2}\right](n_*-n)+\epsilon_c,\quad
\phi_n=\left[\frac{F_1}{c_1}-\frac{F_2}{c_2}\right](n_*-n)+\epsilon_c\frac{\beta}{\alpha},\quad n<n_*
\end{gathered}
\label{eq:StaticSolutionBroken}
\end{equation}

\begin{figure}[!ht]
\minipage{0.5\textwidth}
\center{\includegraphics[width=\linewidth] {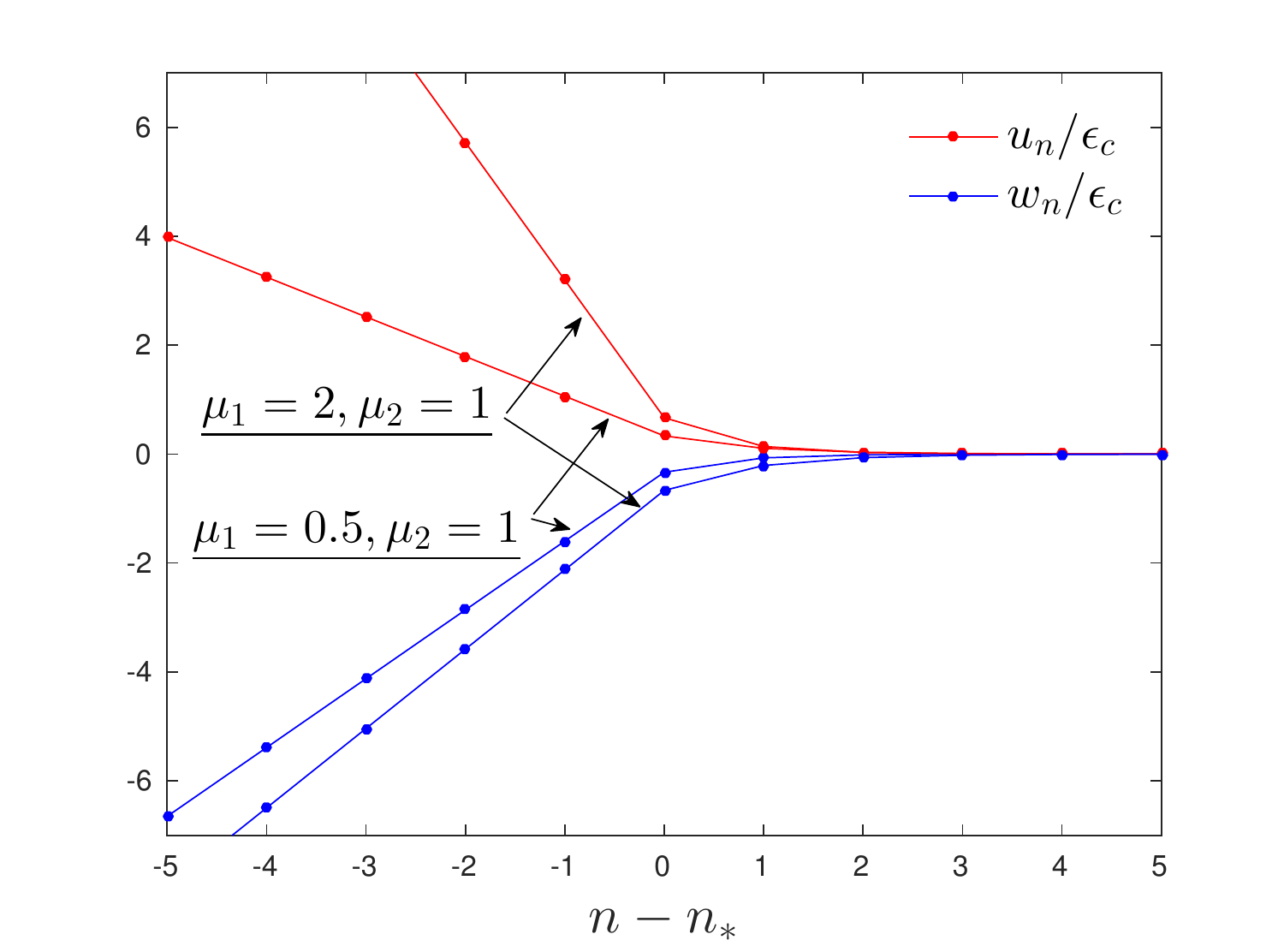} \\ a)}
\endminipage
\hfill
\minipage{0.5\textwidth}
\center{\includegraphics[width=\linewidth]{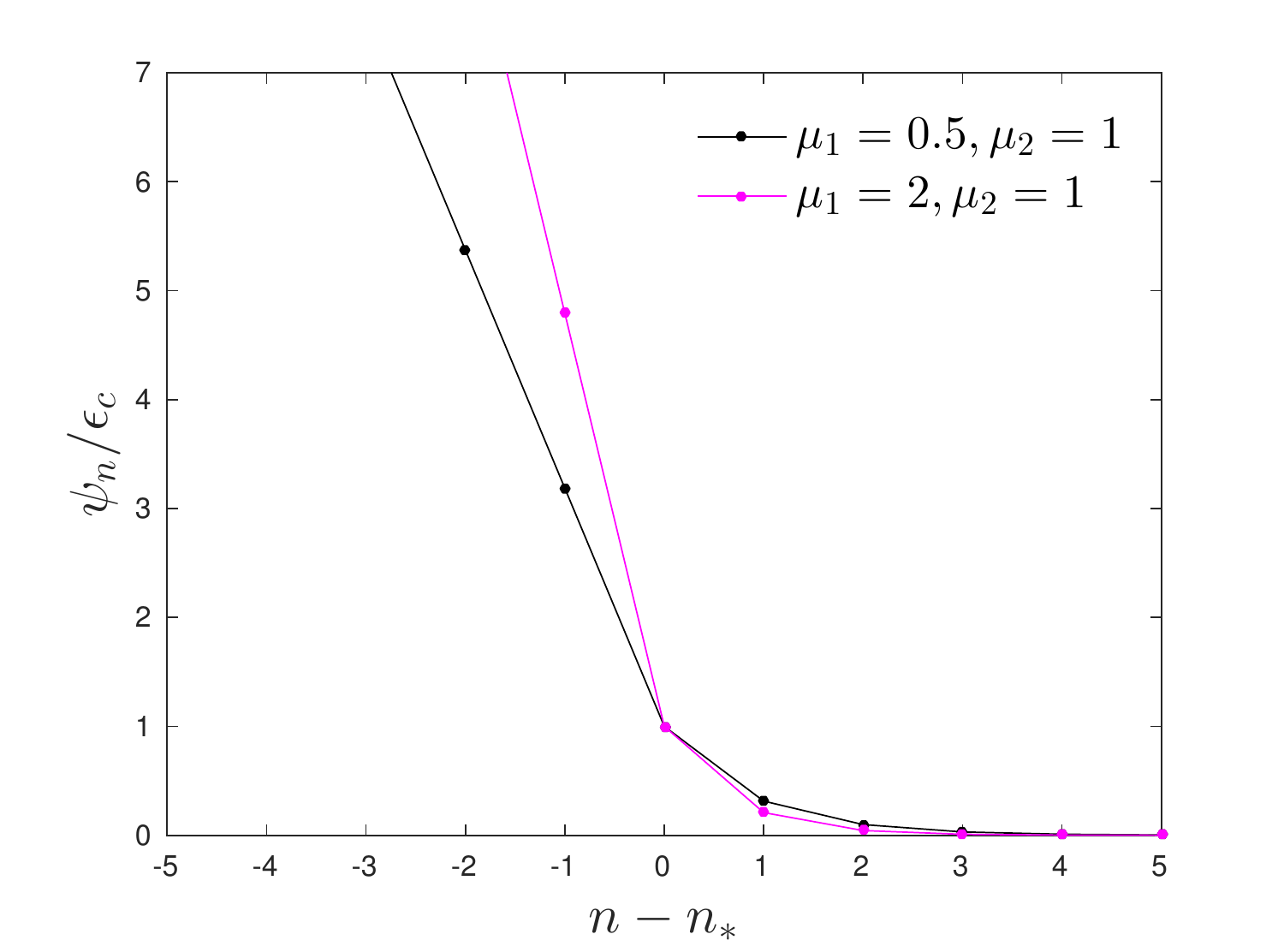} \\ b)}
\endminipage
\caption[ ]{a) Displacements of the chains for two sets of parameters, b) corresponding function $\psi_n$ for the same sets of parameters.}
\label{fig:StaticDisplacements}
\end{figure}

Although the formal solution is derived, we still need to obtain the relations between the forces that cause the fracture and material properties. For that we  consider the equations at $n=n_*$ which in terms of $\psi_n$ and $\phi_n$ are:
\begin{equation*}
\begin{gathered}
(\psi_{n_*-1}-\psi_{n_*})+(\psi_{n_*+1}-\psi_{n_*})-\alpha\psi_{n_*}=0,\\
(\phi_{n_*-1}-\phi_{n_*})+(\phi_{n_*+1}-\phi_{n_*})-\beta\psi_{n_*}=0
\end{gathered}
\end{equation*}
Utilising the derived solutions in \eqref{eq:StaticSolutionIntact}, \eqref{eq:StaticSolutionBroken} and the governing equation for $\lambda$ we arrive to:
\begin{equation*}
\frac{F_1}{c_1}+\frac{F_2}{c_2}=\psi_0\left(\frac{1}{\lambda}-1\right),\quad
\frac{F_1}{c_1}-\frac{F_2}{c_2}=\psi_0\frac{\beta}{\alpha}\left(\frac{1}{\lambda}-1\right).
\end{equation*}
The last set of equations finally allows to determine the dependences of forces:
\begin{equation}
F_1=F_2=F,\quad
\frac{F}{F_0}=\frac{\sqrt{4+\alpha}+\sqrt{\alpha}}{2\sqrt{\alpha}},\quad
F_0=c\epsilon_c.
\label{eq:StaticForce}
\end{equation}

Here, $F_0$ is a static force required to break the structure compounded from the two masses connected by a spring of stiffness $c$. So, the relations \eqref{eq:StaticPsiPhi}, \eqref{eq:StaticSolutionIntact}, \eqref{eq:StaticLambda}, \eqref{eq:StaticSolutionBroken} and \eqref{eq:StaticForce} solve the problem. The displacements for two particular cases are shown in Fig.~\ref{fig:StaticDisplacements}.

Global energy release rate $G$ can be computed by consideration of change in potential energy of the structure when the crack advances by a unite value length $a$. It is enough to consider the change in potential energy while a crack progresses, which is a difference between the work of forces on the change of displacements and elastic energies in the broken part of a structure. The energy release rate $G$, then, can be computed from the following equation:
\begin{equation*}
G=\frac{1}{2a}F_1(u_{n_*-1}-u_{n_*})-\frac{1}{2a}F_2(w_{n_*-1}-w_{n_*})=\frac{F_1^2}{2ac_1}+\frac{F_2^2}{2ac_2}
\end{equation*}
The substitution of \eqref{eq:StaticForce} into the last equation results in:
\begin{equation}
G=\frac{G_0}{\alpha}\left(\frac{1}{\lambda}-1\right)^2,\quad
G_0=\frac{\epsilon_c^2c}{2a}.
\label{eq:StaticERR}
\end{equation}

Quantity $G_0$ is the local energy release rate which is equal to the amount of the released energy when the failure of structure of two masses and spring of stiffness $c$ happens related to a cell of size $a$. This quantity is adjacent two the one for the force $F_0$.

\begin{figure}[!ht]
\minipage{0.5\textwidth}
\center{\includegraphics[width=\linewidth] {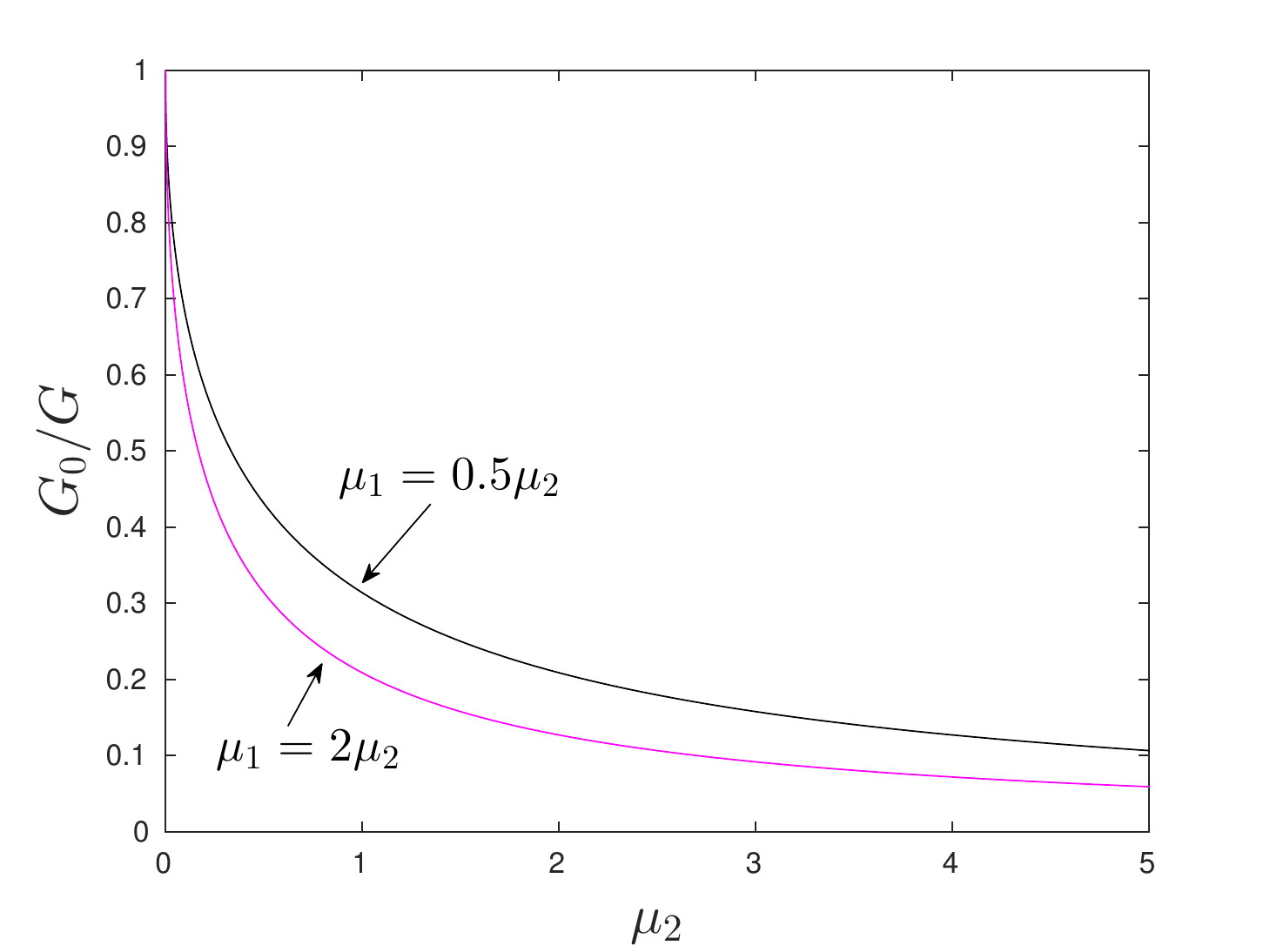} \\ a)}
\endminipage
\hfill
\minipage{0.5\textwidth}
\center{\includegraphics[width=\linewidth]{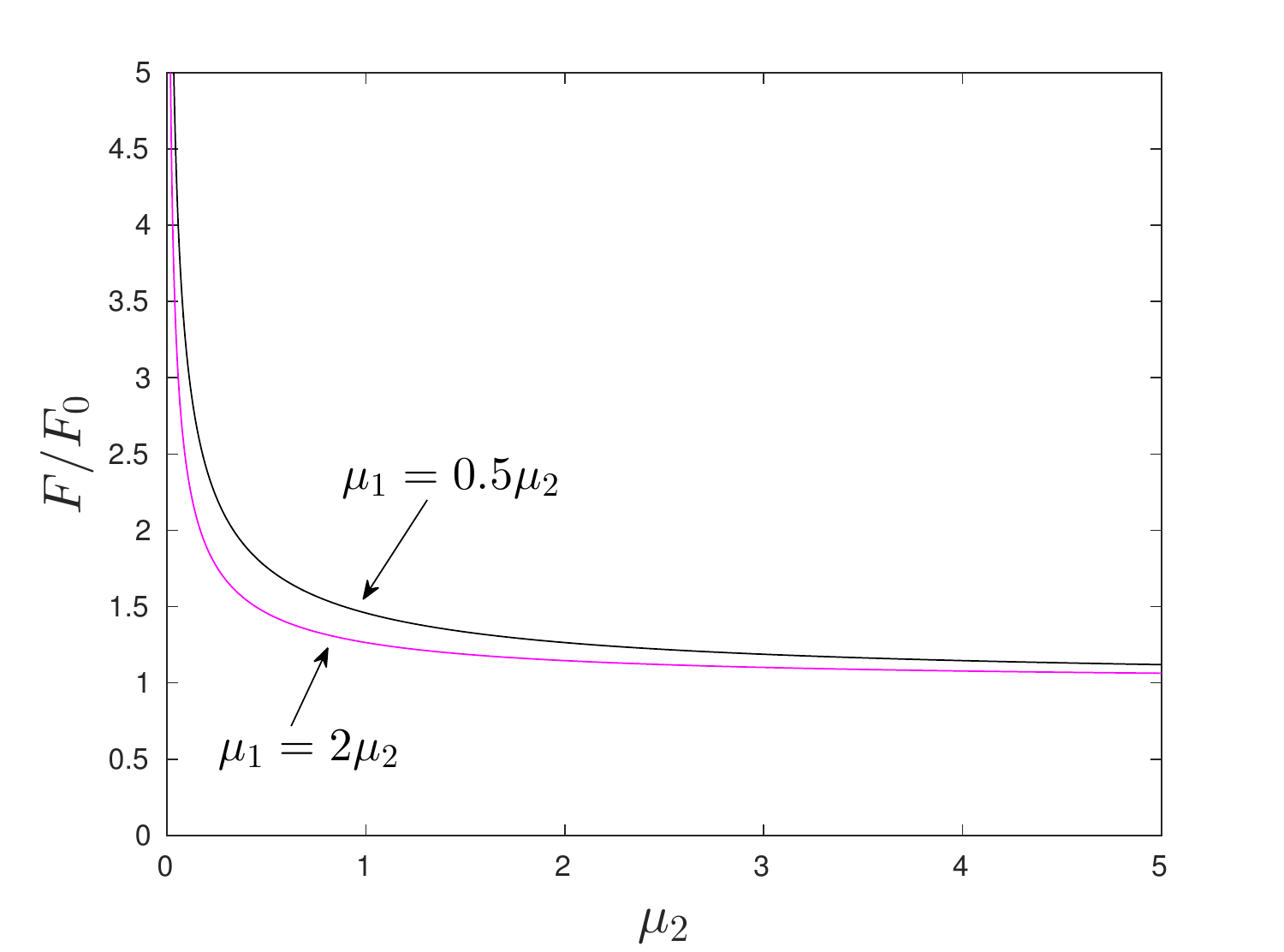} \\ b)}
\endminipage
\caption[ ]{a) Energy release rate ratio $G_0/G$ for two sets of the parameters, b) force ratio $F/F_0$ for the same sets of parameters.}
\label{fig:StaticERR}
\end{figure}

The plots for the ratios $G_0/G$ and $F/F_0$ are presented in Fig.~\ref{fig:StaticERR} for two sets of material parameters. The attention should be paid to the fact that for any set of parameters, not only those that were depicted, we observe $G_0/G<1$ and $F/F_0>1$ which is known as lattice trapping~\cite{thomson1971}. Particularly, this effect demonstrate that in the force required to break such a discrete structure accedes the force required to break its single element separately from the structure. That is a feature of a discrete model which is not observed in the analysis of continuum solids. It shows that at microlevel analysis of the material a special analysis is required.

Finally, we also would like to underline the following important and rather trivially predicted facts valid
for the static solution. First, both the energy and load ratios, $G_0/G$ and $F/F_0$,
are monotonic functions of their parameters. Secondly, the second condition in
\eqref{fracture_cond_0} is always satisfied. Note also that the absolute value is redundunt here as it simply relates to the sign of the applied forces, $F_j>0$.
And, finally, lets us reiterate again that, for the static problem, the total load has be self-balanced ($F_1=F_2=F$) to guarantee that itself is a trivial conclusion.
With these remarks, we turn to the analysis of transient and possible steady-state crack movement and, consequently, the contribution of inertia.

\section{Dynamic problem for the crack moving with a constant speed}

\subsection{Governing equations}

Let us assume that, starting from some moment of time, the crack moves with a constant speed $v$. Its value is unknown a priori and is
a result of the joint action of two moving forces $F_1$ and $F_2$ applied to the different parts of the destroyed dissimilar chain (see Fig.~\ref{fig:DissimilarChain}).
The forces move along the structure with their speeds $v_1^f$ and $v_2^f$, respectively. Those speeds define how fast a signal from one point can reach another point along the respective part of the broken chain.
As a result, one can observe that if $v_j^f<v_j$, $j=1,2$, the crack speed satisfies the following constrain
\begin{equation}
v\le v_{max}\equiv \max\{v_{1},v_{2}\}.
\label{eq:CrackSpeedLimit_0}
\end{equation}
Let us define two quantities important for the future analysis:
\begin{equation}
v_c=\min\{v_{1},v_{2}\},
\label{eq:CrackSpeedLimit_1}
\end{equation}
\begin{equation}
v_*=\frac{c_1+c_2}{m_1+m_2}=\frac{\beta_2^2}{\beta_1^2+\beta_2^2}v_1^2+\frac{\beta_1^2}{\beta_1^2+\beta_2^2}v_2^2.
\label{eq:CrackSpeedLimit_2}
\end{equation}
One can directly check that the second value defines the speed of wave propagating along the part of the chain which is still intact.

In the case $v_1=v_2$ the three values are equal to each other ($v_c=v_*=v_{max}=v_1$) and it represents the limiting value for the crack propagation speed in the structure $v\le v_c$.
Otherwise, the crack speed can exceed the value $v_c$. In general case, $v_c<v_*<v_{max}$ and, as a result, one can expect an existence of various propagation regimes
depending on the applied load and the material properties of the dissimilar chain.
We discuss those cases in details below always assuming that $v_1\le v_2$. In case of inequality, this means that $v_c=v_1$, $v_{max}=v_2$ and $v_1<v_*<v_2$.

Since the crack moves with a constant speed, $v$, it is reasonable to study possible steady-state regimes.
Let us displacements of the mass $u_n(t)$, $w_n(t)$ as functions of a new variable:
\begin{equation}
\eta=n-n_*(t)=n-n_0-vt,
\label{eq:Eta}
\end{equation}
where $n_0$ is a distance between the origin of laboratory coordinate system and a moving frame at time.
Following \cite{slepyan2012}, we assume that the variable $\eta$ represent a continuous, not a discrete,  variable.
Consequently, we require that the displacement of the oscillators be expressed as:
\begin{equation}
u_n(t)=u(\eta,t),\quad w_n(t)=w(\eta,t).
\label{eq:NewFunctions}
\end{equation}
In this dynamic case it is also convenient to consider the same linear combinations of the displacements. Thus, we define:
\begin{equation}
\psi(\eta,t)=u(\eta,t)-w(\eta,t), \quad \phi(\eta,t)=u(\eta,t)+w(\eta,t).
\label{eq:Functions_psi_phi}
\end{equation}

Function $\psi(\eta,t)$ defines a crack opening at the broken part of the structure, $\eta<0$, and it equals to an elongation of a spring between 2 chains in the intact part of the structure, $\eta>0$. The other function demonstrates how the middle line of the structure changes in time. Both pairs of functions, $u(\eta,t)$, $w(\eta,t)$ and $\psi(\eta,t)$, $\phi(\eta,t)$,
do not represent the displacements in the structure but rather the respective traces of the mass movement in time.

The stated fracture criterion \eqref{fracture_cond_0} becomes:
\begin{equation}
\psi(0,t)=\epsilon_c,\quad |\psi(\eta,t)|<\epsilon_c,\quad \eta>0.
\label{eq:FractureCondition_Eta_t}
\end{equation}
Note that we have eliminated the absolute value requirement here. Indeed, adopting the general criterion \eqref{fracture_cond_0}$_1$ (or equivalently $ |\psi(0,t)|=\epsilon_c$)
one may expect that $\psi(0,t)=\pm \epsilon_c$ at different breakage time. This however, contradicts to the assumption of steady-state movement: that at each moment when the link brakes the deformation of the structure (at least locally) remains the same.
Note that other regular movements are also proved to be possible (see for example alternating regime in \cite{mishuris2009localised} or clustering \cite{slepyan2010crack} and forerunning \cite{nieves2016analysis,slepyan2015forerunning}).

Since solution of the dynamic problem can, due to the waves propagating ahead of the crack tip, generally speaking oscillate at the infinity and thus may not have a limiting value there.
This creates some difficulty to recognise when the structure "stays in balance" in time.
We impose the following additional conditions at infinity to avoid possible rigid body motion (compare \eqref{eq:Rigid_body}):
\begin{equation}
\frac{\partial}{\partial t}\phi(\eta,t)=0,\quad t\to\infty,
\label{eq:FractureCondition_Phi_Eta_t}
\end{equation}
and
\begin{equation}
|\psi(\eta,t)|<\infty,\quad|\phi(\eta,t)|<\infty,\quad \eta>0,\quad t\to\infty.
\label{eq:Condition_infinity}
\end{equation}

For a constant crack speed, the equations of motion \eqref{eq:OriginalProblem} become:
\begin{equation}
\begin{gathered}
m_1\left(\frac{\partial^2}{\partial t^2}-2v\frac{\partial^2}{\partial t\partial \eta}+v^2\frac{\partial^2}{\partial\eta^2}\right)u(\eta,t)=\\[2mm]
c_1(u(\eta+1,t)+u(\eta-1,t)-2u(\eta),t)-c(u(\eta,t)-w(\eta,t))H(\eta)+F_1\delta(\eta+n_2+(v-v_1^f)t),\\[2mm]
m_2\left(\frac{\partial^2}{\partial t^2}-2v\frac{\partial^2}{\partial t\partial \eta}+v^2\frac{\partial^2}{\partial\eta^2}\right)w(\eta,t)=\\[2mm]
c_2(w(\eta+1,t)+w(\eta-1,t)-2w(\eta),t)+c(u(\eta,t)-w(\eta,t))H(\eta)-F_2\delta(\eta+n_1+(v-v_2^f)t),
\end{gathered}
\label{eq:OriginalProblemSteadyState}
\end{equation}
where $H(x)$ is a Heaviside step function, $\delta(x)$ is Dirac delta-function and $n_{j},j=1,2$ are the distances between the crack tip and top and bottom forces, respectively, at the beginning of crack motion represented by a fixed speed, $v$.

The consequent application of the Fourier transform and Laplace transform (with the terms from initial conditions being omitted as irrelevant in a limiting case $t\to\infty$):
\begin{equation}
\begin{gathered}
\left[(s+ikv)^2+\omega_1^2(k)\right]U(k,s)=-\beta_1^2\Psi^+(k,s)+\frac{F_1}{m_1}\frac{e^{-ikn_1}}{s+ik(v-v_1^f)},\\
\left[(s+ikv)^2+\omega_2^2(k)\right]W(k,s)=\beta_2^2\Psi^+(k,s)-\frac{F_2}{m_2}\frac{e^{-ikn_2}}{s+ik(v-v_2^f)},
\end{gathered}
\label{eq:OriginalProblemFourierLaplace}
\end{equation}
where
\begin{equation}
\omega_{j}^2(k)=4v_j^2\sin^2\left(\frac{k}{2}\right),\quad j=1,2,
\label{eq:DispersionRelations}
\end{equation}
determine the dispersion relations of the separated chains and are responsible for the wave characteristics of the destroyed part of the structure.
Here, the following standard notations have been introduced:
\begin{equation}
\begin{gathered}
U(k,s)=\int_0^{\infty}\left(\int_{-\infty}^{\infty}u(\eta,t)e^{ik\eta}\,d\eta\right)e^{-st}\,dt,\quad W(k,s)=\int_0^{\infty}\left(\int_{-\infty}^{\infty}w(\eta,t)e^{ik\eta}\,d\eta\right)e^{-st}\,dt,\\[2mm]
\Psi(k,s)=U(k,s)-W(k,s)=\Psi^+(k,s)+\Psi^-(k,s),\\[2mm]
\Psi^\pm(k,s)=\int_0^{\infty}\left(\int_{-\infty}^{\infty}\psi(\eta,t)H(\pm\eta)e^{ik\eta}\,d\eta\right)e^{-st}\,dt.
\end{gathered}
\end{equation}

\subsection{System of Wiener-Hopf equations for a transient regime}

The combination of equations \eqref{eq:OriginalProblemFourierLaplace} reduces problem to a Wiener-Hopf type with respect to the functions $\Psi^\pm(k,s)$:
\begin{equation}
\begin{gathered}
\Psi^-(k,s)+L(k,s)\Psi^+(k,s)=\\[2mm]
\frac{F_1}{im_1(v-v_1^f)}\frac{e^{-ikn_1}}{k-f_1^-}\frac{1}{(s+ikv)^2+\omega_1^2(k)}+\frac{F_2}{im_2(v-v_2^f)}\frac{e^{-ikn_2}}{k-f_2^-}\frac{1}{(s+ikv)^2+\omega_2^2(k)},
\end{gathered}
\label{eq:WienerHopfPsi}
\end{equation}
where
\begin{equation}
f_1^-=\frac{is}{v-v_1^f},\quad f_2^-=\frac{is}{v-v_2^f}.
\label{eq:f_j}
\end{equation}

Kernel function $L(k,s)$ is defined as follows:
\begin{equation}
L(k,s)=1+\frac{\beta_1^2}{(s+ikv)^2+\omega_1^2(k)}+\frac{\beta_2^2}{(s+ikv)^2+\omega_2^2(k)}.
\label{eq:Function_L_s}
\end{equation}

Note that function $L(k,s)$ is analytic and not equal to zero as a function of $\eta$-veriable along the real axis for any value $s>0$ and the following conditions can be directly checked:
\begin{equation}
\begin{gathered}
L(k,s)=\overline{L(-k,s)},\quad
|L(k,s)|=|L(-k,s)|,\quad \text{Arg}L(k,s)=-\text{Arg}L(-k,s),\quad k\in\mathbb{R}, \quad s\in\mathbb{R}_+,\\[2mm]
L(k,s)=1+s^{-2}(\beta_1^2+\beta_2^2)+O(k),\quad k\to0,\quad L(k,s)=1+O\left(\frac{1}{k^2}\right),\quad k\to\infty, \quad s\in\mathbb{R}_+.
\end{gathered}
\end{equation}
As a result, the index of this function (winding number) is zero and the function $L(k,s)$ can be factorised for any fixed value od $s>0$ by means of Cauchy-type integral:
\begin{equation}
L(k,s)=L^+(k,s)L^-(k,s),\quad L^{\pm}(k,s)=\exp\left(\pm\frac{1}{2\pi i}\int_{-\infty}^{\infty}\frac{\text{Log}L(\xi,s)}{\xi-k}\,d\xi\right),
\end{equation}
where functions $L^\pm(k,s)$ are analytic in the half-planes $\pm\Im k>0$, respectively, and both tend to unit as $k\to\infty$.

Equation \eqref{eq:WienerHopfPsi} can now be rewritten in an alternative way:
\begin{equation}
\frac{1}{L^-(k,s)}\Psi^-(k,s)+L^+(k,s)\Psi^+(k,s)=B(k,s),
\label{eq:Wiener_Hopf_Fact}
\end{equation}
that makes is directly applicable for application of Wiener-Hopf technique. Note that the right hand side of the Wiener-Hopf equation
\begin{equation}
B(k,s)\equiv \frac{1}{L^-(k,s)}\sum_{j=1}^2\frac{F_j}{im_j(v-v_j^f)}\frac{e^{-ikn_j}}{k-f_j^-}\frac{1}{(s+ikv)^2+\omega_j^2(k)},
\label{eq:Wiener_Hopf_B}
\end{equation}
is a decreasing function along the real $\eta$-axis and thus can be represented as a sum of two
functions $B^+(k,s)$ and $B^-(k,s)$ vanishing as $k\to\pm\infty$
\[
B^\pm(k,s)=\frac{1}{2\pi i}\int_{-\infty}^\infty B(\xi,s)\frac{d\xi}{\xi-k},\quad \pm\Im k>0.
\]
Then the solution of equation \eqref{eq:Wiener_Hopf_Fact} takes the form:
\begin{equation}
\Psi^-(k,s)=L^-(k,s)B^-(k,s),\quad \Psi^+(k,s)=\frac{1}{L^+(k,s)}B^+(k,s).
\label{eq:Wiener_Hopf_Sol}
\end{equation}

Function $\Phi(k,s)$ is defined as:
\begin{equation}
\Phi(k,s)=U(k,s)+W(k,s)=\int_0^{\infty}\left(\int_{-\infty}^{\infty}\phi(\eta,t)e^{ik\eta}\,d\eta\right)e^{-st}\,dt,
\end{equation}
and is found from the following equation:
\begin{equation}
\Phi(k,s)=-M(k,s)\Psi^+(k,s)+E(k,s),
\label{eq:WienerHopfPhi}
\end{equation}
where
\begin{equation}
M(k,s)=\frac{\beta_1^2}{(s+ikv)^2+\omega_1^2(k)}-\frac{\beta_2^2}{(s+ikv)^2+\omega_2^2(k)}.
\label{eq:WienerHopfPhi_M}
\end{equation}
and
\begin{equation}
E(k,s)=\frac{F_1}{im_1(v-v_1^f)}\frac{e^{-ikn_1}}{k-f_1^-}\frac{1}{(s+ikv)^2+\omega_1^2}
-\frac{F_2}{im_2(v-v_2^f)}\frac{e^{-ikn_2}}{k-f_2^-}\frac{1}{(s+ikv)^2+\omega_2^2}.
\label{eq:WienerHopfPhi_Q}
\end{equation}

As a result, knowing the function $\Psi^+(k,s)$ from \eqref{eq:Wiener_Hopf_Sol}$_{2}$, one can immediately find function $\Phi(k,s)$
\begin{equation}
\Phi(k,s)=-M(k,s)\frac{1}{L^+(k,s)}B^+(k,s)+E(k,s),
\label{eq:WienerHopfPhi_1}
\end{equation}
and thus separately determine $U(k,s)$ and $W(k,s)$ that should be inverted to evaluate
the sought for displacements $u(\eta,t)$ and $w(\eta,t)$. This is, however, rather difficult technical task requiring application two times Cauchy integral and then the inversion of  Fourier and Laplace integral transforms.
On the other hand, the main goal of this research is to analyse possible steady state propagation regimes and determine ways how the transient regimes
may approach the former as $t\to\infty$. Here, one can use Abelian type theorems linking limiting behaviour of functions with their images.

\subsection{Wiener-Hopf equations for the steady-state regime}

In this section we analyse possible steady-state regimes achieved at $t\to\infty$. In this case one can reformulate the problem in terms of the Fourier transform only.
All the functions become to be defined by the single variable $k$:
\begin{equation}
\begin{gathered}
U(k)=\lim_{s\to0+}sU(k,s),\quad W(k)=\lim_{s\to0+}sW(k,s),\\[2mm]
\Psi^{\pm}(k)=\lim_{s\to0+}s\Psi^{\pm}(k,s),\quad \Phi^{\pm}(k)=\lim_{s\to0+}s\Phi^{\pm}(k,s),
\end{gathered}
\label{eq:SteadyStateFunctionsSolution}
\end{equation}
that corresponds to
\[
u(\eta)=\lim_{t\to+\infty}u(\eta,t),\quad w(\eta)=\lim_{t\to+\infty}w(\eta,t).
\]
The auxiliary functions
\begin{equation}
\psi(\eta)=u(\eta)-w(\eta),\quad \phi(\eta)=u(\eta)+w(\eta),
\label{eq:SteadyStateFunctionsSolution_origin}
\end{equation}
satisfy the fracture condition \eqref{eq:FractureCondition_Eta_t} that at the steady-state regime takes the form:
\begin{equation}
\psi(0)=\epsilon_c,\quad |\psi(\eta)|<\epsilon_c,\quad \eta>0.
\label{eq:FractureCondition_Eta}
\end{equation}
and the  condition \eqref{eq:Condition_infinity} is useful to write in the form
\begin{equation}
\int_0^\infty\psi(\xi)d\xi<\infty,\quad
\int_0^\infty(\phi(\xi)-\phi_\infty)d\xi<\infty.
\label{eq:BalanceCondition_Phi_Eta_0}
\end{equation}
where the integrals should be considered in sense of distributions while the constant $\phi_\infty$ is unknown a priori and needed to be computed from the solution.
Thus effectively means that the function $\Psi^+(k)$ is bounded near point $k\to0$, while
\begin{equation}
\Phi^+(k)\sim\frac{\phi^+_\infty}{0-ik},\quad k\to0.
\label{eq:BalanceCondition_Phi_asymp_0}
\end{equation}
Note also that function $\phi(\eta)$ (and in  some cases also $\psi(\eta)$) can linearly grow as $\eta\to-\infty$.
In other words:
\begin{equation}
\Psi^-(k)\sim-\frac{\psi^-_\infty}{(0+ik)^2},\quad
\Phi^-(k)\sim-\frac{\phi^-_\infty}{(0+ik)^2},\quad k\to0.
\label{eq:BalanceCondition_Phi_Psi_asymp_0}
\end{equation}
Finally, since the functions $\phi(\eta)$ and $\psi(\eta)$ are continuous at zero, one can expect that:
\begin{equation}
\Psi^\pm(k)\sim \frac{\psi_0}{0\mp ik},\quad
\Phi^\pm(k)\sim \frac{\phi_0}{0\mp ik},\quad k\to\infty.
\label{eq:BalanceCondition_Phi_Psi_asymp_0}
\end{equation}

Now we multiply equations \eqref{eq:Wiener_Hopf_Fact} and
 \eqref{eq:WienerHopfPhi} by $s$ and pass it to limit $s\to0+$ to transform them to
\begin{equation}
\frac{1}{L^-(k)}\Psi^-(k)+L^+(k)\Psi^+(k)=B(k),
\label{eq:Wiener_Hopf_Fact_0}
\end{equation}
\begin{equation}
\Phi(k)=-M(k)\Psi^+(k)+E(k),
\label{eq:WienerHopfPhi_0}
\end{equation}
where
\begin{equation}
\begin{gathered}
L(k)=\lim_{s\to0+}L(k,s),\quad
M(k)=\lim_{s\to0+}M(k,s),\quad L^\pm(k)=\lim_{s\to0+}L^\pm(k,s),
\end{gathered}
\label{eq:Function_L}
\end{equation}
and
\[
B(k)=\lim_{s\to0+}s\, B(k,s),\quad E(k)=\lim_{s\to0+}s\, E(k,s).
\]
One can check the following estimates:
\begin{equation}
L(k)=\frac{\Xi}{k^2}+O(1),\quad k\to0,\quad L(k)=1+O(k^{-2}),\quad k\to\infty,
\label{L_kernel}
\end{equation}
and
\[
\frac{M(k)}{L(k)}=\Upsilon+O(k^2),\quad k\to0,\quad \frac{M(k)}{L(k)}=O(k^{-2}),\quad k\to\infty,
\]
where
\begin{equation}
\Xi=\frac{(\beta_1^2+\beta_2^2)(v_*^2-v^2)}{(v_1^2-v^2)(v_2^2-v^2)},\quad
\Upsilon=\frac{\beta_1^2(v_2^2-v^2)-\beta_2^2(v_1^2-v^2)}{\beta_1^2(v_2^2-v^2)+\beta_2^2(v_1^2-v^2)}.
\label{eq:ConstantK}
\end{equation}

The behaviour of the factors $L^{\pm}(k)$ at infinity can be found as:
\begin{equation}
L^{\pm}(k)=1\pm i\frac{Q}{k}+O\left(\frac{1}{k^2}\right),\quad k\to\infty,
\label{eq:Asymptotics_L+-_infty}
\end{equation}
where the constant $Q=Q(v)$ is computed from the integral:
\[
Q=\frac{1}{\pi}\int_{0}^{\infty}\log{|L(k)|}\,dk,
\]
which converges due to the estimate \eqref{L_kernel} above.

Asymptotics of the factors $L^\pm(k)$ near zero point depends essentially on the value of the crack speed $v$.
Depending on the sign of the parameter $\Xi$ from \eqref{L_kernel}, one can represent the kernel function $L(k)$ in different manners.
Thus if $\Xi=\Theta^2>0$, that is true if $v<v_c$ or $v_*<v<v_{max}$, we can write:
\begin{equation}
L(k)=\frac{(\Theta+ik)(\Theta-ik)}{(0+ik)(0-ik)}L_*(k),\quad L^\pm(k)=\frac{(\Theta\mp ik)}{(0\mp ik)}L_*^\pm(k),
\label{L_kernel_2}
\end{equation}
where $L_*(k)$ tends to unity at zero and infinity. As a result, one can prove that
\begin{equation}
L^{\pm}(k)= \frac{ R^{\pm1}\Theta}{0\mp ik}\left(1+(0\mp ik)(S+1/\Theta)\right)+O(k),\quad k\to0,
\label{eq:Asymptototics_L+-_zero}
\end{equation}
where $\Theta=\Theta(v)$, $R=R(v)$ and $S=S(v)$ should be computed from the relationships:
\begin{equation}
R=\exp\left(\frac{1}{\pi}\int_{0}^{\infty}\frac{\text{Arg}L_*(k)}{k}\,dk\right),
\quad
S=\frac{1}{\pi}\int\limits_0^\infty\frac{\log{|L_*(k)|}}{k^2}\,dk.
\label{eq:R}
\end{equation}
For the remaining case $v_c<v<v_*$, another representation of the kernel $L(k)$ is needed:
\begin{equation}
L(k)=\frac{(\Theta+ik)^2}{(0+ik)^2}L_*(k),\quad L^+(k)=L_*^+(k),\quad L^-(k)=\frac{(\Theta+ik)^2}{(0+ik)^2}L_*^-(k),
\label{L_kernel_3}
\end{equation}
where $\Xi=-\Theta^2<0$ in this case. Then asymptotics of the factors are:
\begin{equation}
L^{+}(k)= R\big(1+(0\mp ik)S\big)+O(k^2),\quad k\to0,
\label{eq:Asymptototics_L+-_zero_3a}
\end{equation}
\begin{equation}
L^{-}(k)= \frac{\Theta^2}{R(0+ ik)^2}\left(1+(0+ ik)(S+2/\Theta)+O(k^2)\right),\quad k\to0,
\label{eq:Asymptototics_L+-_zero_3b}
\end{equation}
where $R$ and $S$ again computed by the formula \eqref{eq:R}.

One can check that the right hand sides $B(k)\equiv 0$ and $E(k)\equiv 0$ for any $|k|>\varepsilon>0$. They represent generalised functions (distributions)
such that:
\begin{equation}
B(k)=b\left(\frac{1}{0+ik}+\frac{1}{0-ik}\right),\quad E(k)=\frac{e_1}{R\,\Theta}\left(\frac{1}{0+ik}+\frac{1}{0-ik}\right)L_-(k)-
\frac{e_2}{R\,\Theta}\left(\frac{1}{0+ik}+\frac{1}{0-ik}\right),
\label{eq:Delta_Dirac}
\end{equation}
where coefficients $b$, $e_1$ and $e_2$ take different values depending on the value of the crack speed $v$ and depends on the loading.

\subsection{Possible steady-state regimes}

\subsubsection{Solutions of the Wiener-Hopf equations}

Solution to the equation \eqref{eq:Wiener_Hopf_Fact_0} can be immediately found
\begin{equation}
\Psi^+(k)=\frac{b}{0-ik}\frac{1}{L^+(k)},\quad \Psi^-(k)=\frac{b}{0+ik}L^-(k).
\label{eq:Psi_solutions}
\end{equation}
Taking into account the asymptotic behaviour of the factors $L^\pm(k)$ given in \eqref{eq:Asymptotics_L+-_infty} and \eqref{eq:Asymptototics_L+-_zero},
we can immediately find the asymptotics of solution at infinity:
\begin{equation}
\Psi^{\pm}(k)=\pm \frac{ib}{k}\left(1-\frac{Qi}{k}\right)+O\left(\frac{1}{k^3}\right),\quad k\to\infty,
\end{equation}
and at zero point:
\begin{equation}
\begin{gathered}
\Psi^-(k)=\frac{b\Theta}{R}\frac{1}{(0+ik)^2}\left(1+(0+ik)\Big(S+1/\Theta\big)\right)+O(1),\quad k\to0,\\[2mm]
\Psi^+(k)=\frac{1}{\Theta}\,\frac{b}{R}+O(k),\quad k\to0.
\end{gathered}
\end{equation}

The fracture condition \eqref{eq:FractureCondition_Eta} determines the relation between the loading and the critical elongation $\epsilon_c$:
\begin{equation}
b(v)=\epsilon_c.
\label{eq:fracture_condition_result}
\end{equation}
As a result, the solution for function $\psi(\eta)=u(\eta)+w(\eta)$ is expressed in terms of the inverse Fourier transform:
\begin{equation}
\begin{gathered}
\psi(\eta)=\frac{\epsilon_c}{2\pi}\int_{-\infty}^{\infty}\frac{1}{0-ik}\frac{1}{L^+(k)},\quad \eta>0,\\[2mm]
\psi(\eta)=\frac{\epsilon_c}{2\pi}\int_{-\infty}^{\infty}\frac{1}{0+ik}L^-(k),\quad \eta<0.
\end{gathered}
\label{eq:Solution_psi}
\end{equation}
and its asymptotics can be computed:
\begin{equation}
\begin{gathered}
\psi(\eta)=\epsilon_c(1-Q\eta)+O(\eta^2),\quad \eta\to0,\\[2mm]
\psi(\eta)=\frac{\epsilon_c}{R}\,\frac{1}{\Theta}\,\left(-\eta+\Big(S+1/\Theta)\big)\right)+O(1),\quad \eta\to-\infty,\\[2mm]
\psi(\eta)=O(1),\quad \eta\to\infty.
\end{gathered}
\end{equation}

Let us now find the other auxiliary function $\phi(\eta)$ representing the sum of the displacement \eqref{eq:SteadyStateFunctionsSolution_origin}.
Equation \eqref{eq:WienerHopfPhi} can be conveniently rewritten using
\eqref{eq:Psi_solutions} and \eqref{eq:fracture_condition_result}
in the form:

\begin{equation}
\Phi(k)=-M(k)\frac{\epsilon_c}{0-ik}\frac{1}{L^+(k)}+
\frac{e_1}{0+ik}\left(\frac{1}{0+ik}+\frac{1}{0-ik}\right)-
e_2\left(\frac{1}{0+ik}+\frac{1}{0-ik}\right),
\label{eq:WienerHopfPhi_00}
\end{equation}
or
\begin{equation}
\Phi(k)=-\frac{M(k)}{L(k)}\frac{\epsilon_c}{0-ik}L^-(k)+
\frac{e_1}{0+ik}\left(\frac{1}{0+ik}+\frac{1}{0-ik}\right)-
e_2\left(\frac{1}{0+ik}+\frac{1}{0-ik}\right).
\end{equation}

One can show that the condition
\begin{equation}
e_1(v)=\frac{\epsilon_c\Upsilon(v)}{\Theta(v) R(v)},
\label{eq:second_condition}
\end{equation}
guarantees that the plus function $\Phi^+(k)$ is bounded near zero point while the minus function $\Phi^-(k)$ is singular at zero point.
One can find the sought for function $\phi(\eta)$ by directly applying inverse Fourier transform.

\subsubsection{Analysis of possible regimes}

\emph{\textbf{Subsonic regime}} ($v<v_c$)

In this case conditions \eqref{eq:fracture_condition_result} and \eqref{eq:second_condition}
we can conclude that the steady state regime preserving the position of the dissimilar chain in space can be achieved if the forces $F_1$ and $F_2$
applied to the different parts of the chain are interrelated throw the following equation:
\begin{equation}
\frac{F_j}{F_0}= \frac{\Theta(v)}{R(v)}\frac{v_j-v_j^f}{v_j-v},\quad j=1,2, \quad v<v_c.
\label{eq:ForcesVsCrackSpeed_less_vc}
\end{equation}
where $F_0$ is the critical static force defined in \eqref{eq:StaticForce}$_2$.

Note that in the case when $v_1=v_2=v_c$ we recover the condition obtained for the homogeneous chain:
\[
\frac{F_1}{v_c-v_1^f}=\frac{F_2}{v_c-v_2^f},
\]
and if the forces are moving with the same velocity ($v_1^f=v_2^f$) then, for the balance condition to be satisfied, we recover the static balance condition (i.e. that the forces are the same). Thus, in case of dynamic fracture of dissimilar chain to preserve the system in balance, the acting forces applied from different parts of the chain are not equal and should depend on the achieved crack speed, $v$.

\vspace{2mm}
\noindent
\emph{\textbf{Intersonic regime}} ($v_1<v<v_2$)
In this case it turns out that force $F_1$ does not effect the fracture process at all and the only applied force $F_2$ affects possible crack movement.
Moreover, there is a gap in the crack velocity, $v_c<v<v_*$, when the crack cannot propagate at all. Consider those cases in a bit more details.

\vspace{2mm}
\noindent
\underline{{\it First intersonic region} ($v_1<v<v_*$).} Here no solution exists as neither from the conditions
\eqref{eq:fracture_condition_result} or \eqref{eq:second_condition} can be satisfied.

\vspace{2mm}
\noindent
\underline{{\it Second intersonic region} ($v_*<v<v_2$).}

\begin{equation}
\frac{F_2}{F_0}=\frac{v_2-v_2^f}{R(v)\,\Xi(v)}\,\frac{v_2+v}{\mu_2v_2^2}\,\frac{(v-v_1^2)(v_2^2-v_*^2)}{(v-v_*^2)(v_2^2-v_1^2)},\quad v_*<v<v_2.
\label{eq:ForcesVsCrackSpeed_greater_vc}
\end{equation}

In the last expression the integral parameter $R$ is given by the previous formula and quantity $F_0$ is shown in \eqref{eq:StaticForce}. The last expressions turn out to be easy to verify by the numerical simulation of a dynamical system which is demonstrated below.

\subsubsection{Numerical examples for displacement profiles}

\begin{figure}[!ht]
\minipage{0.5\textwidth}
\center{\includegraphics[width=\linewidth] {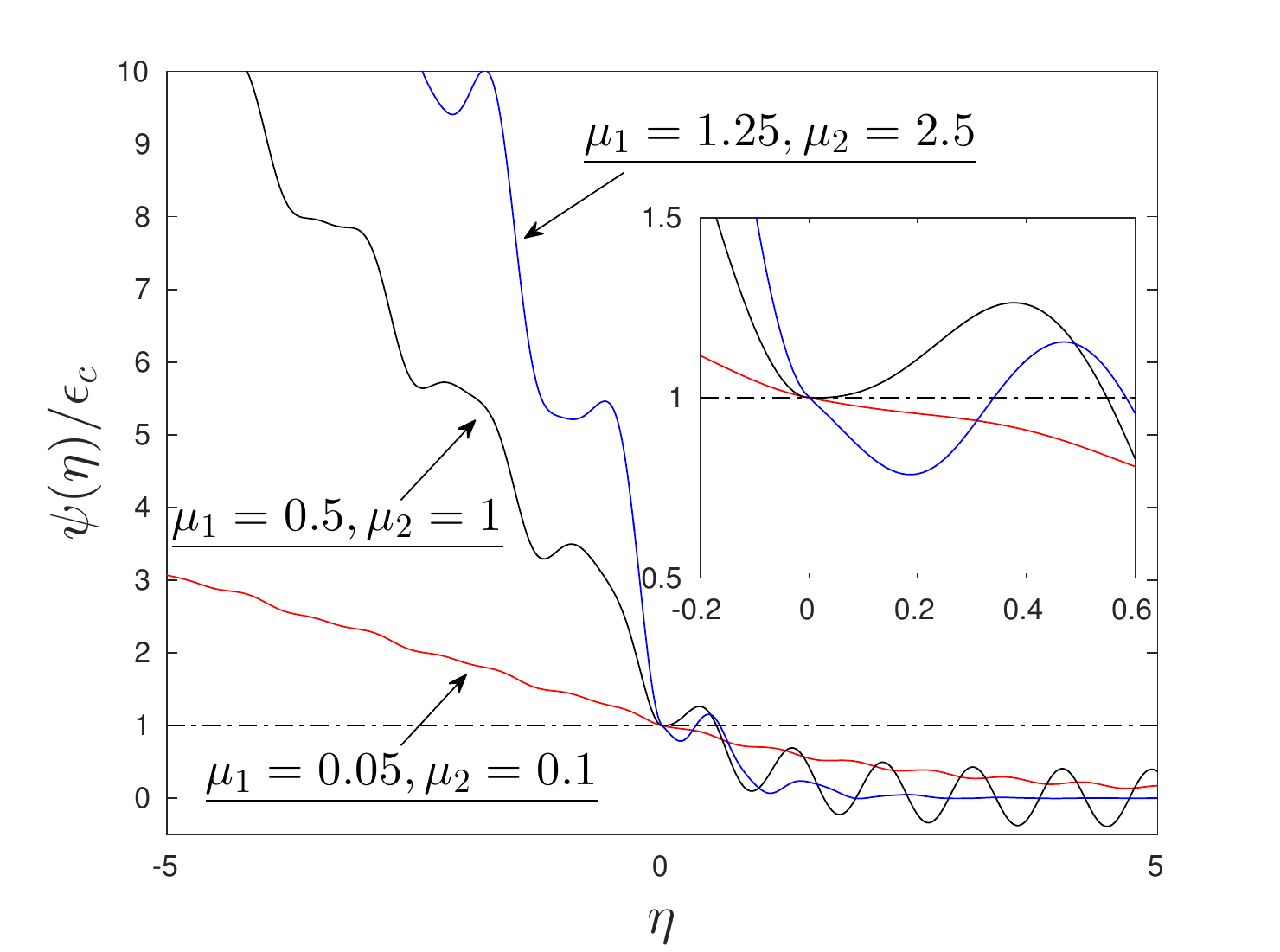} \\ a)}
\endminipage
\hfill
\minipage{0.5\textwidth}
\center{\includegraphics[width=\linewidth]{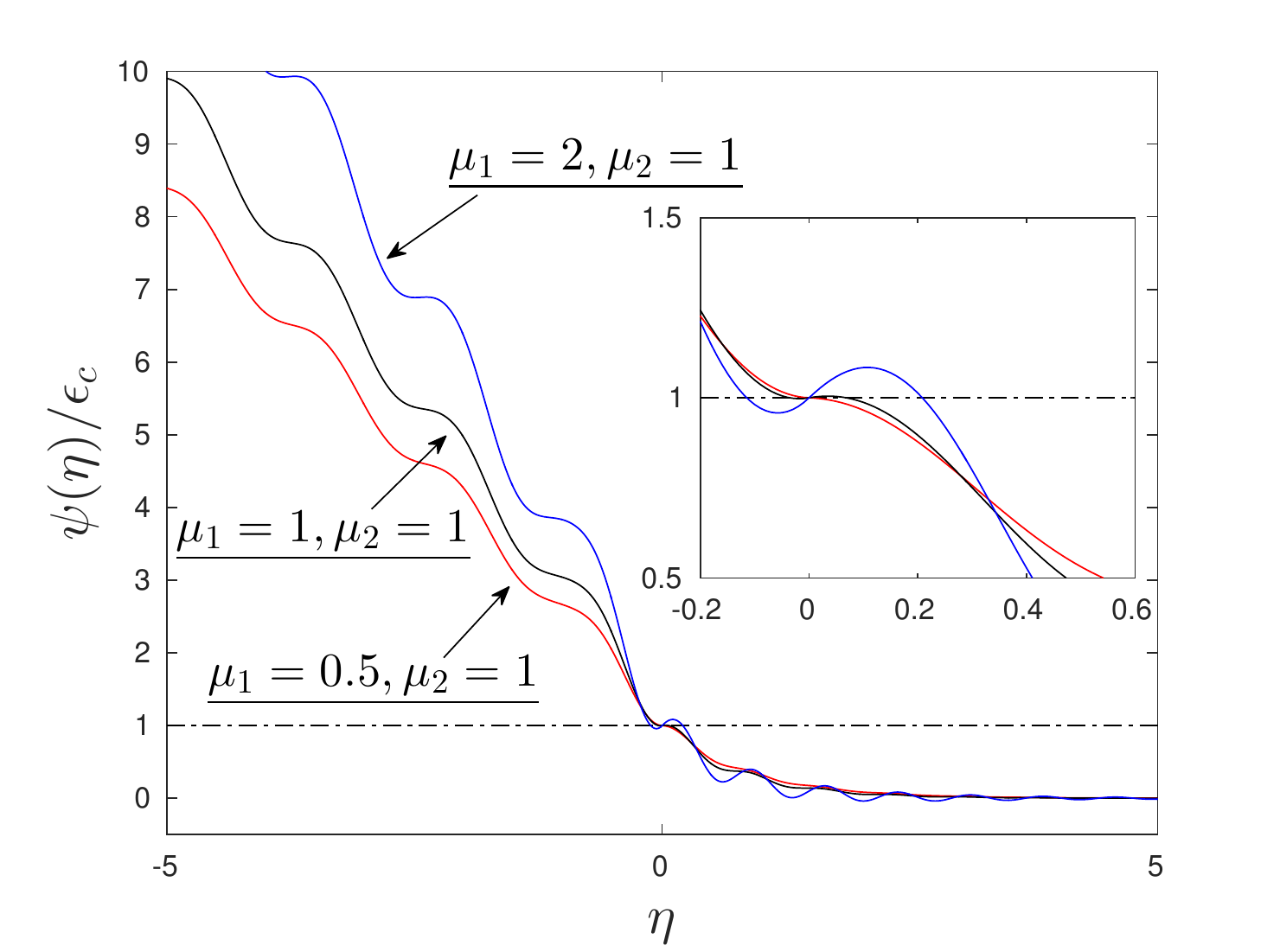} \\ b)}
\endminipage
\caption[ ]{Function $\psi(\eta)$ for different values of model parameters: a) $\mu_2=2\mu_1,v_1=v_2,v=0.2v_c$, b) $\mu_2=1,v_1=v_2,v=0.3v_c$. The inserts show magnified plots at the vicinity of a crack tip.}
\label{fig:Psi}
\end{figure}

\begin{figure}[!ht]
\minipage{0.5\textwidth}
\center{\includegraphics[width=\linewidth] {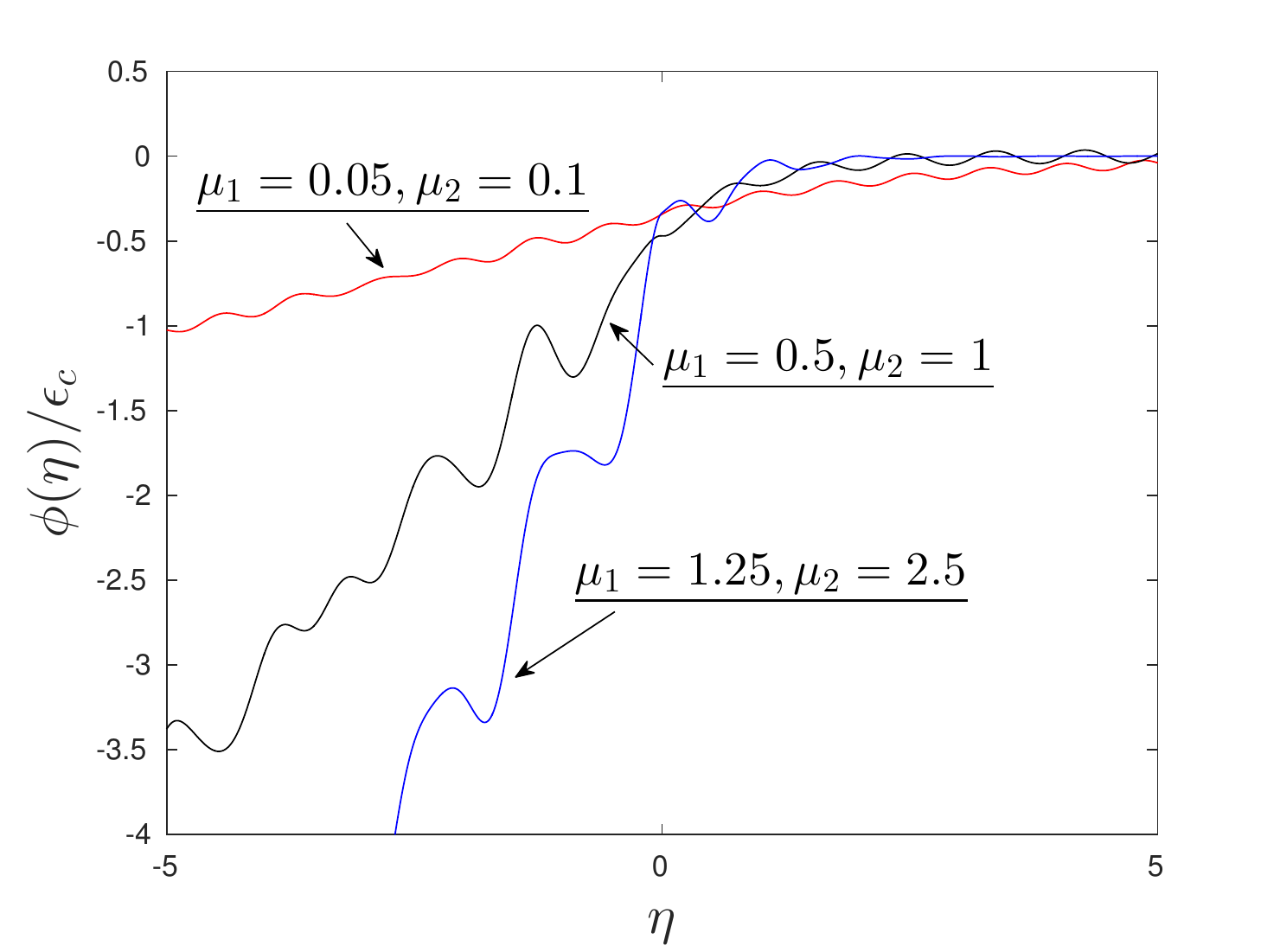} \\ a)}
\endminipage
\hfill
\minipage{0.5\textwidth}
\center{\includegraphics[width=\linewidth]{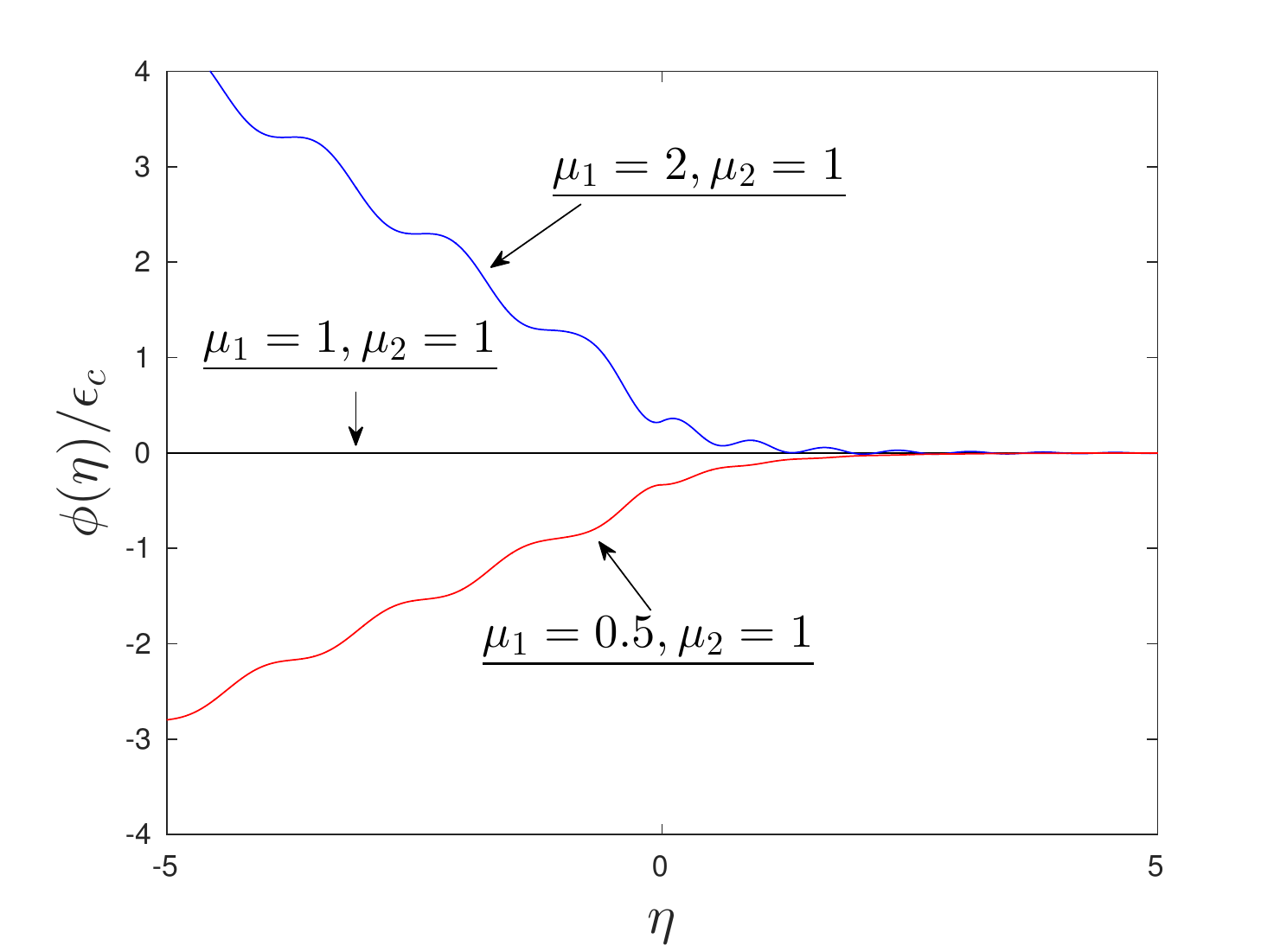} \\ b)}
\endminipage
\caption[ ]{Function $\phi(\eta)$ for different values of model parameters: a) $\mu_2=2\mu_1,v_1=v_2,v=0.2v_c$, b) $\mu_2=1,v_1=v_2,v=0.3v_c$.}
\label{fig:Phi}
\end{figure}

Recall that function $\psi(\eta)$ expresses the crack opening in the broken part of the structure and the elongation of springs between the chains in the intact part of the double chain. For that reason its investigation is important. The evaluation of this function, according to equation \eqref{eq:Solution_psi}, requires firstly setting of certain values of model parameters and a crack speed as well. The plot of $\psi(\eta)$ for several cases is shown in Fig.~\ref{fig:Psi}. Together with it we show the results for $\phi(\eta)$ in Fig.~\ref{fig:Phi} for the same settings.

The feature which is worth of mentioning is that one can see the waves of different lengths emanating from a crack tip both in Fig.~\ref{fig:Psi} and Fig.~\ref{fig:Phi}. This peculiarity is common for fracture problems of discrete media. This is an essential trait of these problems in comparison with the dynamic fracture of continuum media.

The evaluation of function $\psi(\eta)$ is important for checking a second part in the criterion \eqref{eq:FractureCondition_Eta} which states the unique location of the crack tip. The violation of such criterion leads can be vividly observed in Fig.\ref{fig:Psi}a) for $\mu_1=0.5,\mu_2=1$ and $\mu_1=1.25,\mu_2=2.5$, as well as in Fig.\ref{fig:Psi}b) for $\mu_1=1,\mu=1$ and $\mu_1=2,\mu=1$. The examination of different crack speeds and sets of parameters allow to distinguish two different sets of solutions. We call:
\begin{itemize}
\item a fracture regime \textbf{admissible} for a certain set of model parameters and a crack speed if criterion \eqref{eq:FractureCondition_Eta} is completely fulfilled,
\item otherwise such regime \textbf{forbidden}.
\end{itemize}

The further demonstration of the results become easier when the energetic properties of the fracture process are considered.

At the same time condition \eqref{eq:FractureCondition_Phi_Eta_t} is still has to be satisfied. The derivation of a final form of the solution requires detailed analysis of the equations and the functions which is moved to the appendix.

\section{Energy versus crack speed diagram}

\begin{figure}[!ht]
\minipage{0.5\textwidth}
\center{\includegraphics[width=\linewidth] {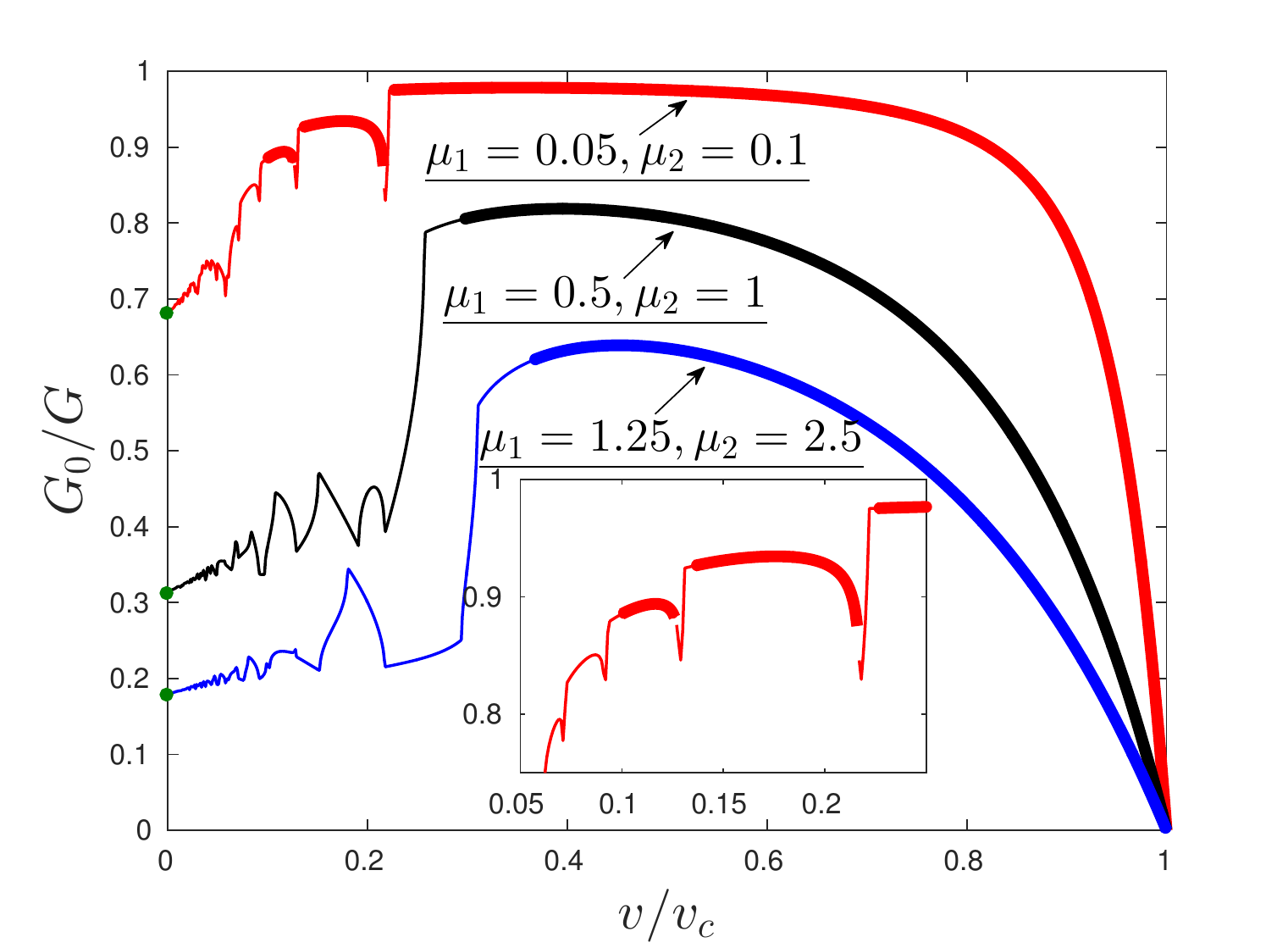} \\ a)}
\endminipage
\hfill
\minipage{0.5\textwidth}
\center{\includegraphics[width=\linewidth]{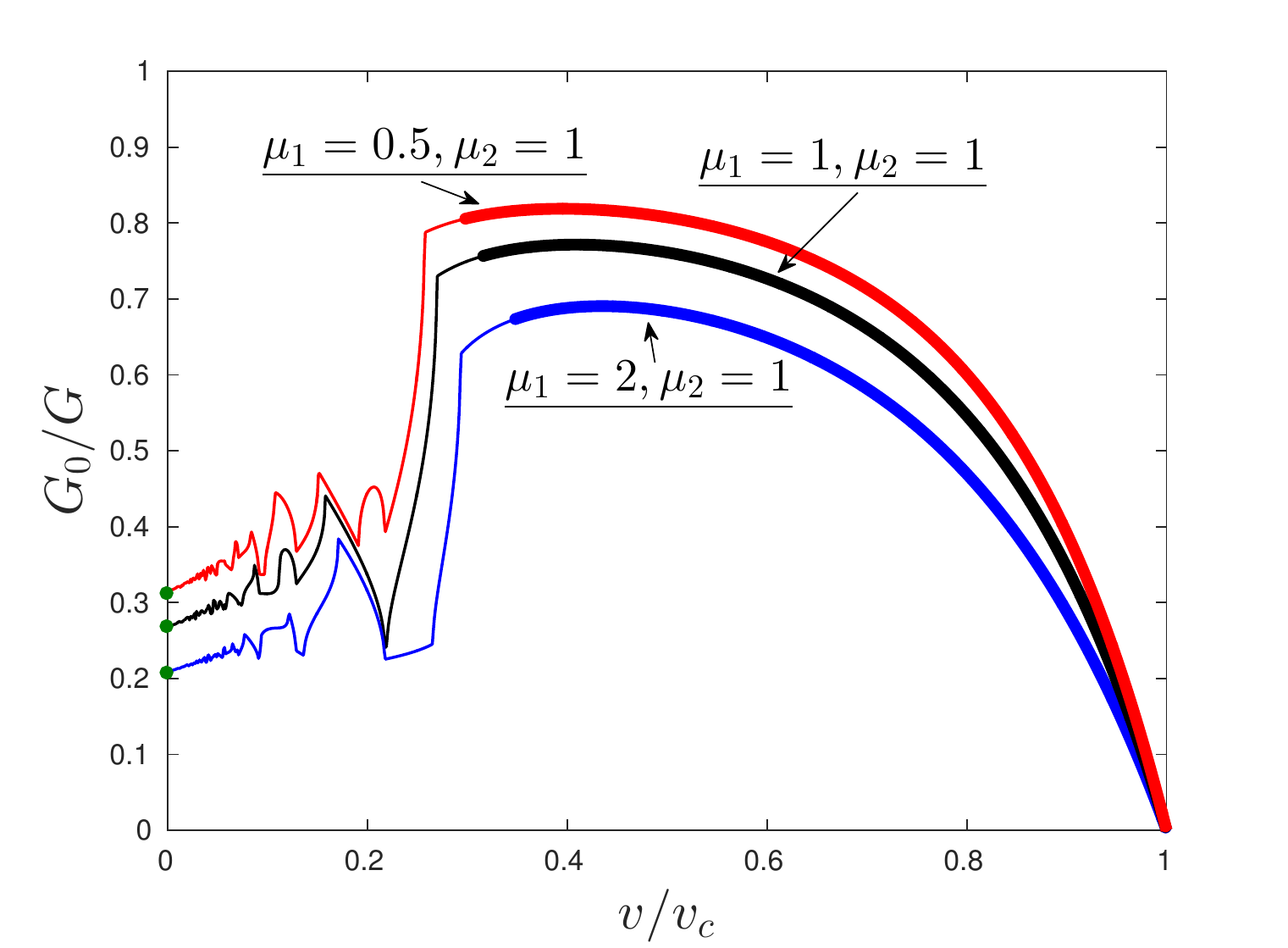} \\ b)}
\endminipage
\caption[ ]{Energy release rates ratio $G_0/G$ for different sets of parameters under a condition $v_1=v_2$: a) $\mu_2=2\mu_1$, b) $\mu_2=1$. Admissible regimes -- thick lines, forbidden regimes -- normal lines, green markers stand for the limiting values when $v\to0$ given by \eqref{eq:StaticERR}.}
\label{fig:ERR_v1=v2}
\end{figure}

Following the same notations used for a static problem, we would like to explore the effect of model parameters on the energy release rate and admissible regimes. The energy release ratio in this case is determined by the following relation~\cite{slepyan2012}:
\begin{equation}
\frac{G_0}{G}=R^2,
\label{eq:ERR_Ratio}
\end{equation}
where quantity $G_0$ is defined in \eqref{eq:StaticERR} and function $R$ is given in \eqref{eq:R}. This ratio demonstrates the released energy carried out by the elastic waves seen in Fig.\ref{fig:Psi} in comparison with the fracture energy. In other words, the lesser this ratio is the more energy is contained in elastic waves. Several examples of these dependences are displayed in Fig.\ref{fig:ERR_v1=v2}. For these examples we set a constraint on the parameters $v_1=v_2$ which allows to reduces the choices of parameters. We also marked the limiting values of $G_0/G$ for $v\to0$ computed by \eqref{eq:StaticERR}. Moreover, we already demonstrate the performed analysis of admissible and forbidden regimes in this plots.

Fig.\ref{fig:ERR_v1=v2}a) refers to the situation when the stiffness of vertical springs $c$ was varied. The interesting point is that the range of admissible regimes grow with the decrease of $c$. For instance, for case $\mu_1=0.05,\mu_2=0.1$ there are three distinct intervals of admissible and forbidden regimes. Such particularity is also observed for a high contrast in model parameters for a simple chain and a square-cell lattice.

The other plot, Fig.\ref{fig:ERR_v1=v2}b), demonstrate the fact that even though the macroscopic properties remain the same, such as speeds of sound $v_{1,2}$ and stiffness $c$ of the springs between the chains, the microlevel fracture properties vary. The energy release rates take different values and, additionally, the qualitative changes of admissible regimes happen. We see the growth of admissible regime with the increase in $\mu_1$ while $\mu_2$ remains the same for all presented cases. Furthermore, for a chosen value of a crack speed there are quantitative changes that are reflected in Fig.\ref{fig:Psi}b) for $\psi(\eta)$, where we see the difference in the values of this function and difference in the wavelengths of the radiated waves.

\begin{figure}[!ht]
\minipage{0.5\textwidth}
\center{\includegraphics[width=\linewidth] {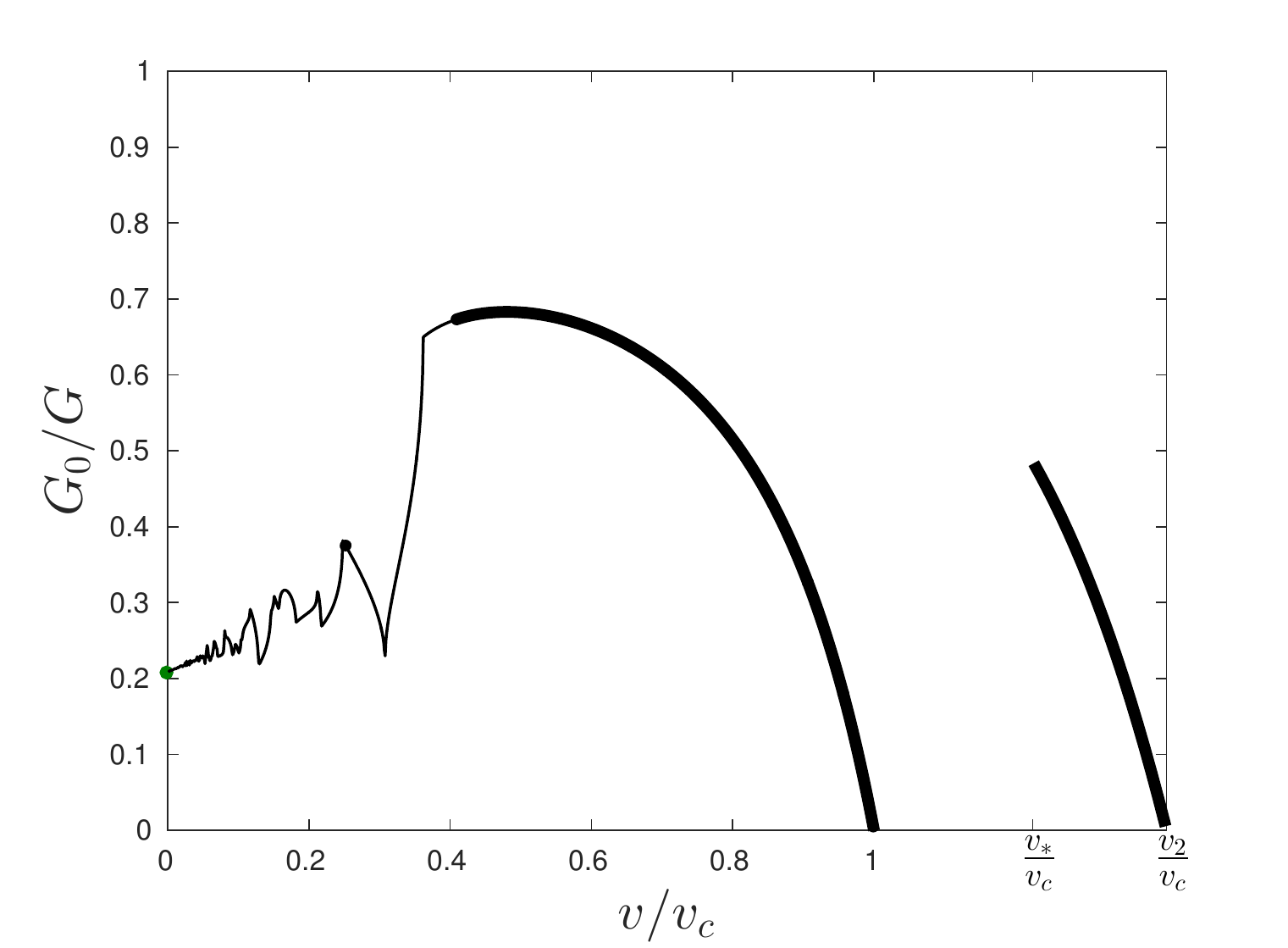} \\ a)}
\endminipage
\hfill
\minipage{0.5\textwidth}
\center{\includegraphics[width=\linewidth]{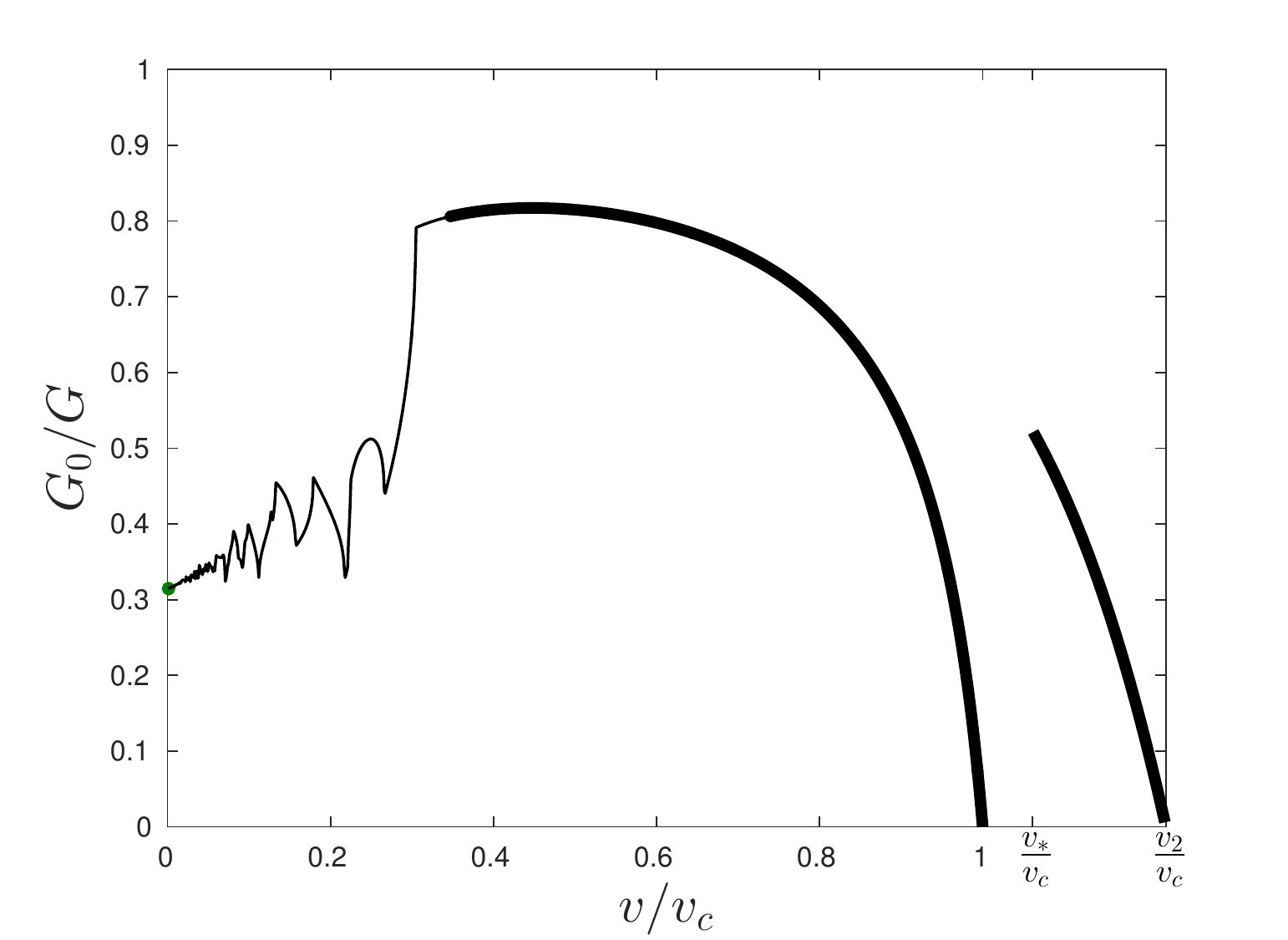} \\ b)}
\endminipage
\caption[ ]{Energy release rates ratio $G_0/G$ for different sets of parameters: a) $\mu_1=2,\mu_2=1,v_1^2/v_2^2=0.5$, b) $\mu_1=0.5,\mu_2=1,v_1^2/v_2^2=2/3$. Admissible regimes -- thick lines, forbidden regimes -- normal lines, green markers stand for the limiting values when $v\to0$ given by \eqref{eq:StaticERR}.}
\label{fig:ERR_diff_v_c}
\end{figure}

In the previous plots we studied the cases when we had a constraint in parameters $v_1=v_2$. The different situation occurs when these parameters are different. In Fig.\ref{fig:ERR_diff_v_c} we display the examples when the speeds of sound are different in two chain. In Fig.\ref{fig:ERR_diff_v_c}a) the parameters are chosen in such a way that $v_1^2=0.5v_2^2$ whereas in Fig.\ref{fig:ERR_diff_v_c}b) we have $v_1^2=2/3v_2^2$. For both cases $v_2$ is the same but according to \eqref{eq:SpeedsOfSound} we have different values of $v_c$. We also provide the analysis of admissible regimes and limiting value at $v\to0$ according to \eqref{eq:StaticERR}.

It is interesting to notice that the ratio $G_0/G$ takes different values for the presented results. We can also observe that for the lower value of $v_1$ in Fig.\ref{fig:ERR_diff_v_c}a) we achieve lower values of $G_0/G$ in comparison with the results in  Fig.\ref{fig:ERR_diff_v_c}b). The attention on these plots is attracted by the intervals of $v$ that correspond to the values $v>v_c$. There is a vivid restriction of the whole range of possible values of the crack speed by the values $\min{(v_1,v_2)},v_*$ and $\max{(v_1,v_2)}$.

Although all the demonstrated results for $G_0/G$ reveal interesting feature of fracture in the structure under consideration we would also like to investigate the dependences of the force. One minor drawback of $G_0/G$ plots is that they are not monotonic even for the intervals of admissible regimes. This, in turn, leads to a non-uniqueness of determination of an achieved steady-state crack speed which is supported by different studies~\cite{ayzenberg2014,marder1995,mishuris2009,pechenik2002,slepyan2012}. Moreover, derived relations for $F_1$ and $F_2$ in \eqref{eq:ForceSpeeds} allow easier verification of delivered solution by numerical simulations. Finally, it is able to show the effects of different model parameters on the admissible regimes in terms of the applied load.

\section{Force versus crack speed diagram}

In this section we analyse the results obtained for the relations of the forces in \eqref{eq:ForceSpeeds}. We also performed the numerical simulations by solving equations \eqref{eq:OriginalProblem} for a finite number of oscillators with free ends boundary conditions. We used 3000 masses of each sort giving total of 6000 masses. The forces are located 10 masses apart from the original crack tip in order to reach the steady-state faster. In the computations we applied the forces calculated for a certain crack speed by \eqref{eq:ForceSpeeds} and recorded the instantaneous crack speed as a function of fracture time. The instantaneous crack speed eventually stabilises and oscillates around a certain value which, by averaging the corresponding data points, was taken as a steady-state crack speed. For that we allowed 1000 fracture events to happen. The settings of the simulations are similar to the ones in~\cite{gorbushin2017dynamic}, where the details are discussed. Thus, we may check the validity of \eqref{eq:ForcesVsCrackSpeed_less_vc} and \eqref{eq:ForcesVsCrackSpeed_greater_vc}.

\begin{figure}[!ht]
\minipage{0.5\textwidth}
\center{\includegraphics[width=\linewidth] {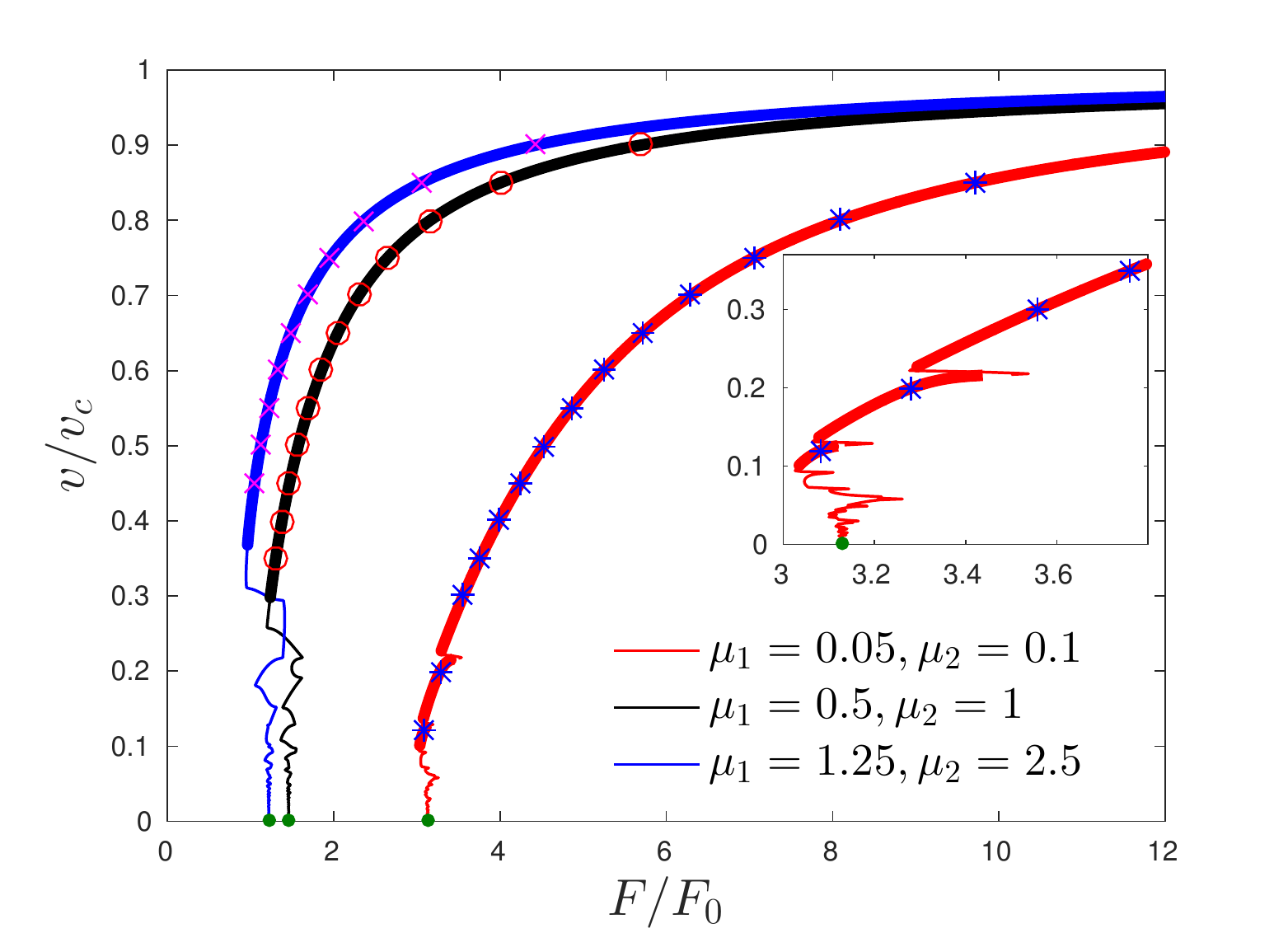} \\ a)}
\endminipage
\hfill
\minipage{0.5\textwidth}
\center{\includegraphics[width=\linewidth]{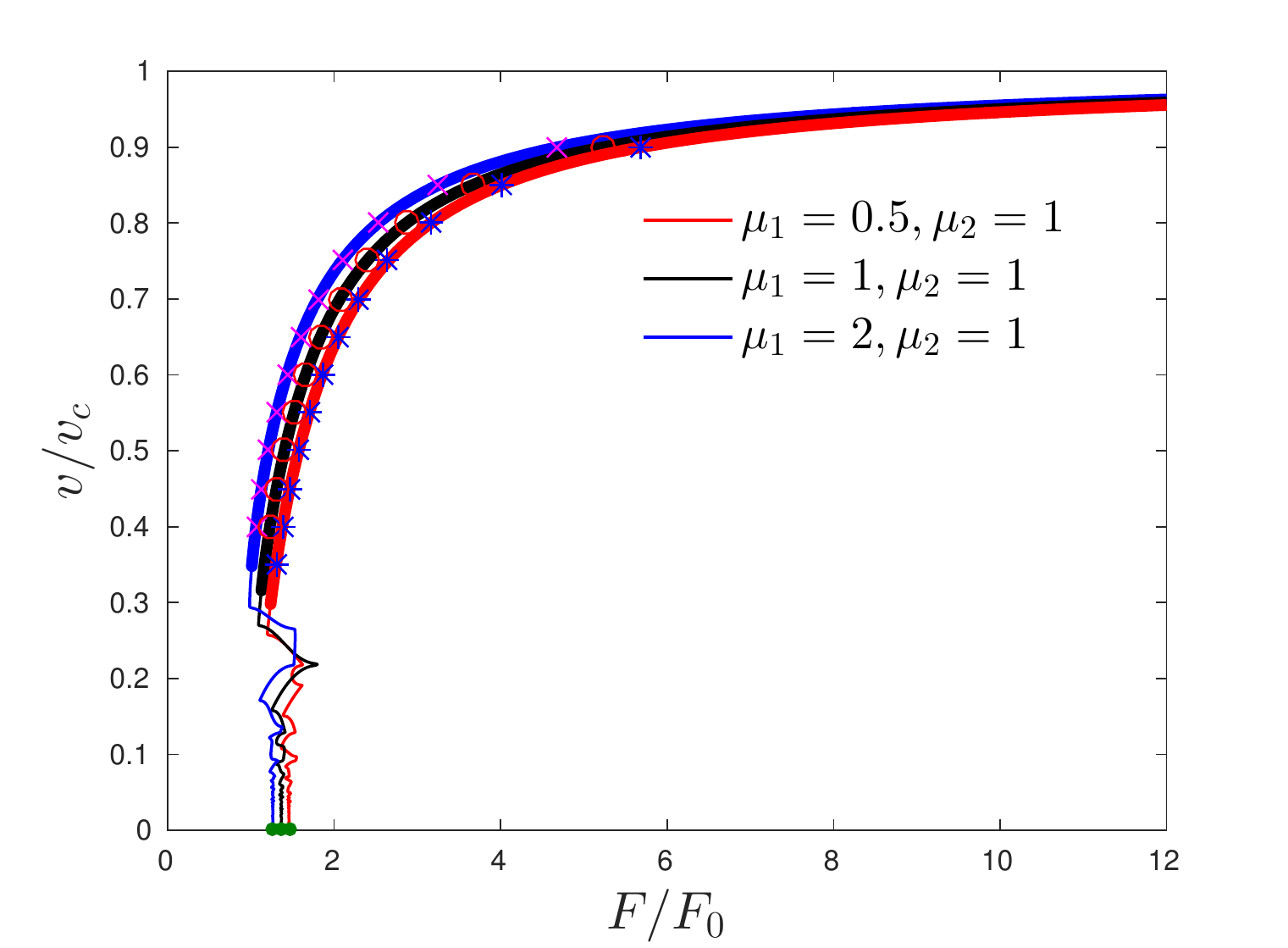} \\ b)}
\endminipage
\caption[ ]{Dependence of normalised force $F/F_0$, according to \eqref{eq:ForceSpeeds} and \eqref{eq:Force_Speed_plot}, for different sets of parameters under a condition $v_1=v_2$ and $v_1^f=v_2^f=0$: a) $\mu_2=2\mu_1$, b) $\mu_2=1$. Admissible regimes -- thick lines, forbidden regimes -- normal lines, green markers stand for the limiting values when $v\to0$ given by \eqref{eq:StaticForce}. The other markers demonstrate the results of numerical simulation after solving dynamical system of equations \eqref{eq:OriginalProblem}.}
\label{fig:Force_v1=v2}
\end{figure}

All the presented results are given for the cases of a fixed force. The effect of different values of a force speed is shown in~\cite{gorbushin2017dynamic}. Firstly, we start with the presentations of the results for cases of $v_1=v_2$. From formula \eqref{eq:ForcesVsCrackSpeed_less_vc} it follows that:
\begin{equation}
F_1=F_2=F,\quad v_1=v_2,\quad v_1^f=v_2^f=0,
\label{eq:Force_Speed_plot}
\end{equation}
where $F_0$ is defined in \eqref{eq:StaticForce}. The dependences of the force ratio $F/F_0$ are plotted in Fig.\ref{fig:Force_v1=v2}. These plots complement those for the energy release ratio in Fig.\ref{fig:ERR_v1=v2}. The limiting values for $v\to0$ are computed with \eqref{eq:StaticForce} and different markers show the results from the numerical simulation of \eqref{eq:OriginalProblem} as described above.

The important feature of the plots $F/F_0$ is that they provide monotonic correlation between $v$ and $F$ within admissible regimes, besides one case $\mu_1=1.25,\mu_2=2.5$ in Fig.\ref{fig:ERR_v1=v2}a).

\begin{figure}[!ht]
\minipage{0.5\textwidth}
\center{\includegraphics[width=\linewidth] {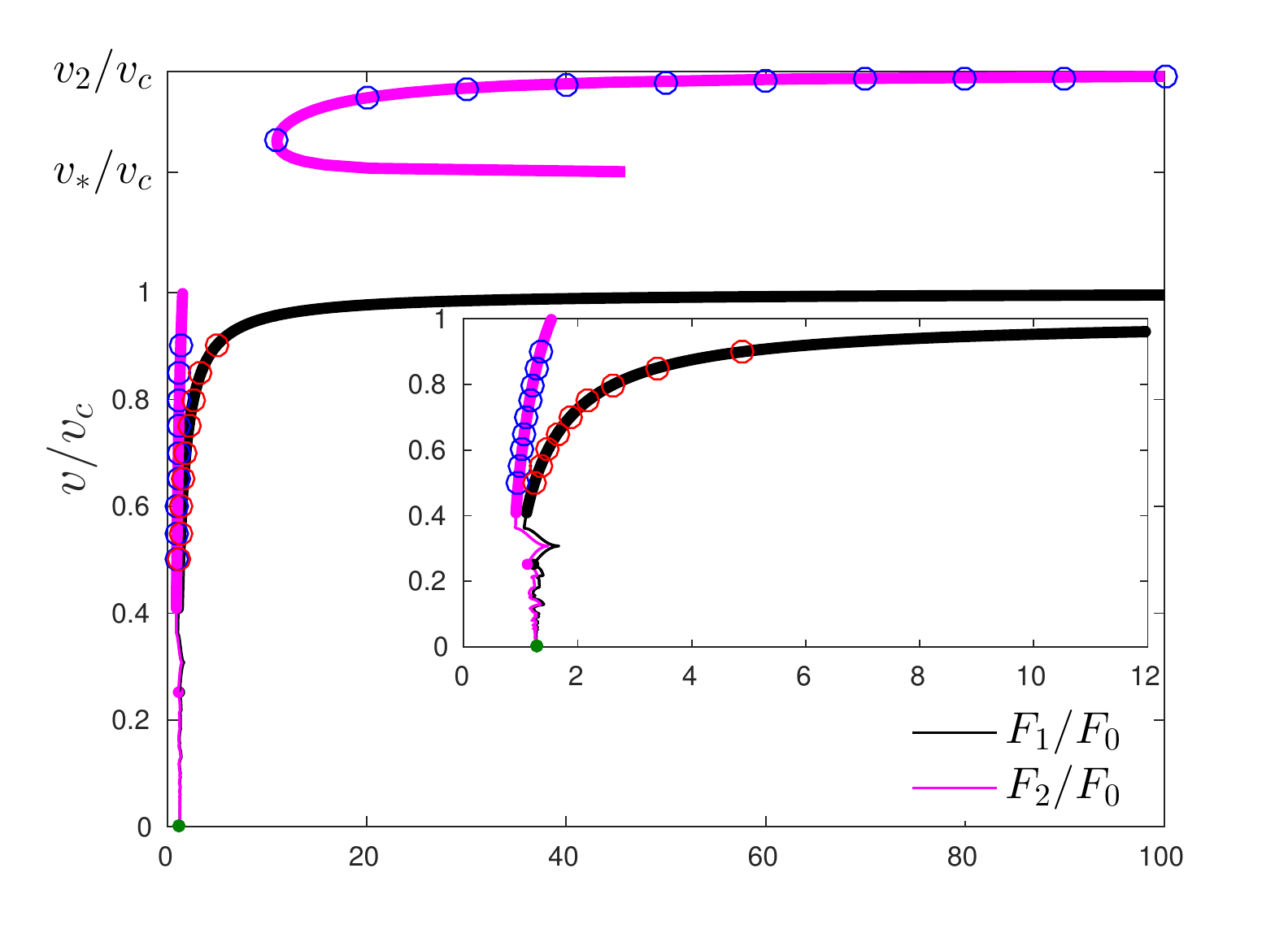} \\ a)}
\endminipage
\hfill
\minipage{0.5\textwidth}
\center{\includegraphics[width=\linewidth]{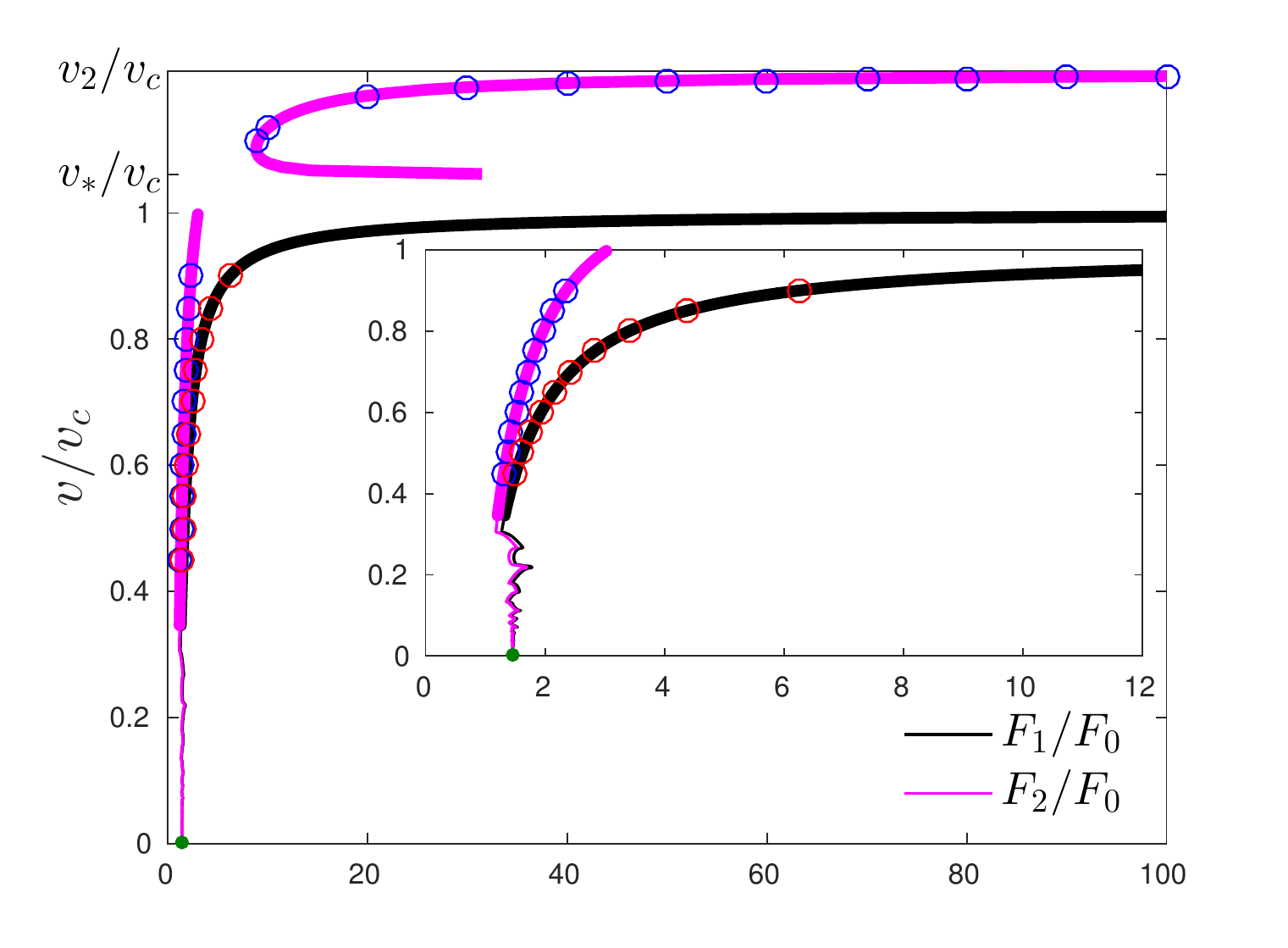} \\ b)}
\endminipage
\caption[ ]{Dependence of normalised forces $F_1/F_0$ and $F_2/F_0$, according to \eqref{eq:Force_Speed_plot}, for different sets of parameters and and $v_1^f=v_2^f=0$: a) $\mu_1=2,\mu_2=1,v_1^2/v_2^2=0.5$, b) $\mu_1=0.5,\mu_2=1,v_1^2/v_2^2=2/3$. Admissible regimes -- thick lines, forbidden regimes -- normal lines, green markers stand for the limiting values when $v\to0$ given by \eqref{eq:StaticForce}. The other markers demonstrate the results of numerical simulation after solving dynamical system of equations \eqref{eq:OriginalProblem}.}
\label{fig:Force_diff_v_c}
\end{figure}

The interesting aspect that is not captured by $G_0/G$ for the cases of different $v_1$ and $v_2$ in Fig.~\ref{fig:ERR_diff_v_c} is that the values of applied forces $F_1$ and $F_2$ should be chosen differently. This evidence is displayed in Fig.~\ref{fig:Force_diff_v_c}. Notice, that if the forces are chosen in a random way, i.e. not belonging to the curves in Fig.~\ref{fig:Force_diff_v_c}, there is a possible stable crack propagation. But in such cases the displacement have a linear growth in time.

\begin{figure}[!ht]
\minipage{0.5\textwidth}
\center{\includegraphics[width=\linewidth] {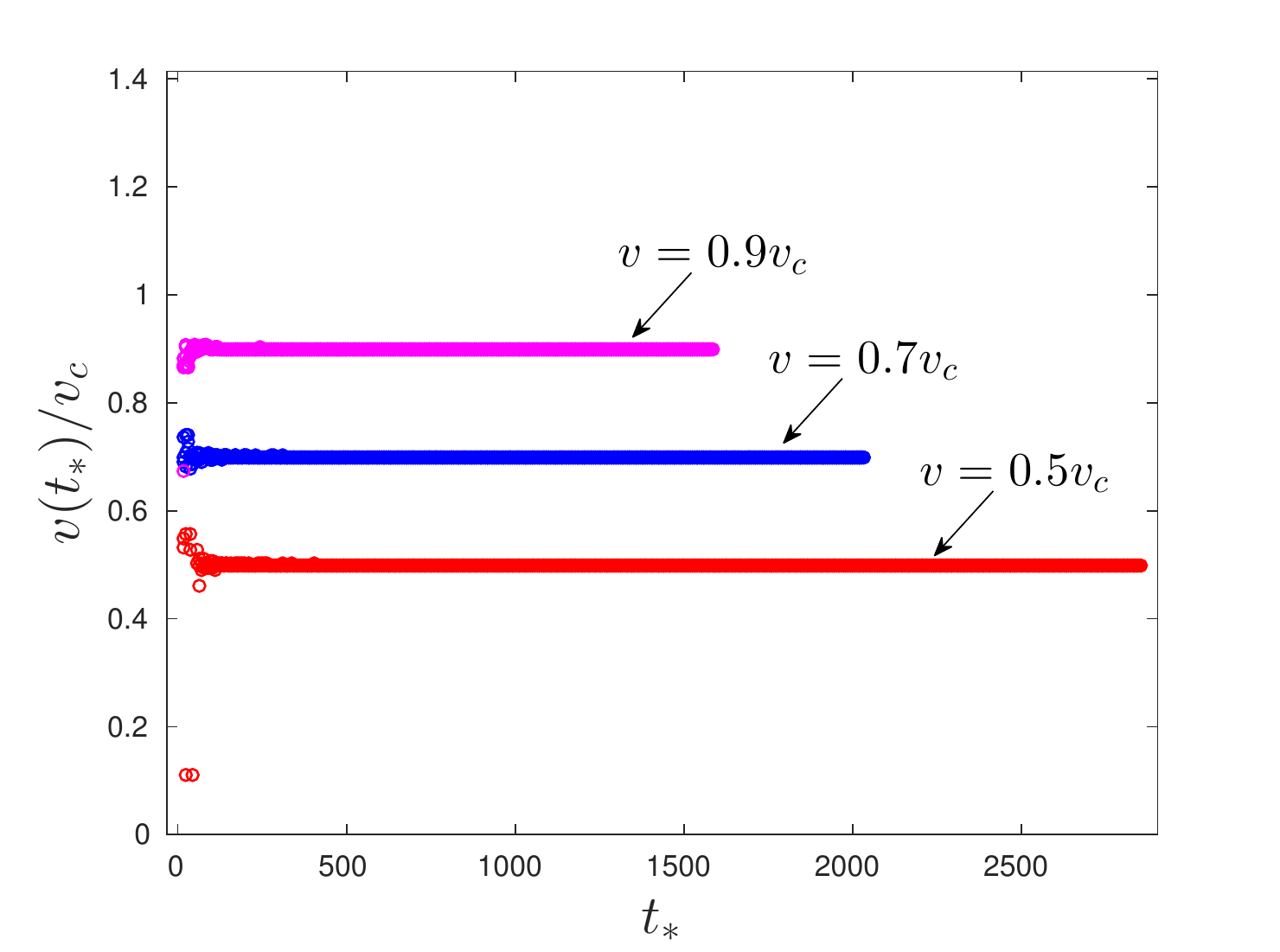} \\ a)}
\endminipage
\hfill
\minipage{0.5\textwidth}
\center{\includegraphics[width=\linewidth]{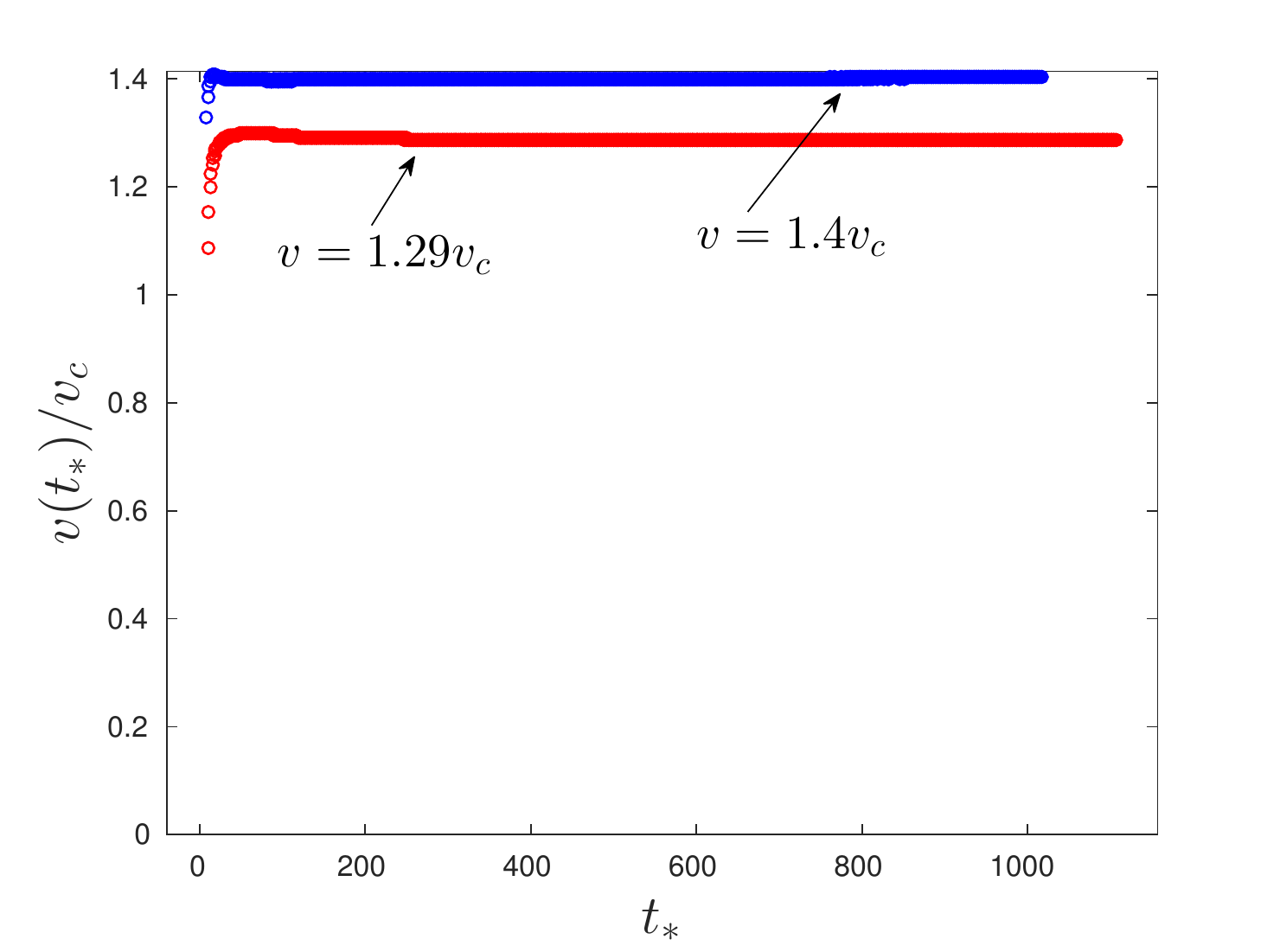} \\ b)}
\endminipage
\caption[ ]{Examples of instantaneous crack speeds for the settings $\mu_1=2,\mu_2=1,v_1^2/v_2^2=0.5$: a) subsonic cases, b) intersonic cases. The indicated values $v$ correspond to the steady-state crack speeds.}
\label{fig:InstantSpeed}
\end{figure}

After performed numerical simulations no crack speeds in a range $v_c<v<v_*$ have been observed. At the same time, for the crack speeds $v>v_*$ there was know difference in obtained values with a change of force $F_1$. This
confirms the fact that when $v_1<v_2$ the main role in fracture plays force $F_2$
making the other force irrelevant. Indeed, the signal from the force $F_1$, which travels along the chain with a lower speed of sound, is not able to reach the crack tip that moves faster than $v_1$, i.e. $v>v_1$.

The final remark concerns the achievement of the steady-state crack speeds. The numerical simulations of dynamic system of equations allows to record the instantaneous crack speed. Examples of the data for the case of $\mu_1=2,\mu_2=1$ and $v_1^2/v_2^2=0.5$ are shown in Fig.~\ref{fig:InstantSpeed}. On that picture $v(t_*)$ is the crack speed calculated at fracture time $t_*$ by means of forward finite differences. The choice of the forces that lead to subsonic speeds is shown in Fig.~\ref{fig:InstantSpeed}a) whereas the intersonic speeds are displayed in Fig.~\ref{fig:InstantSpeed}b). As long as the original distance between the crack tip and forces location was small enough, the instantaneous crack speed stabilizes relatively fast to the predicted steady-state value. Notice that oscillations around the established value of the crack speed are very small both in subsonic and intersonic cases.

\section{Conclusion}
It is well known that the fracture process in bi-material solid can reveal the effects that are not observed with monomaterials. However, most of theoretical works are concerned with continuum media. This work discusses the effects of mismatch in material parameters that occur in discrete structures. Particularly, we studied the steady-state separation of a double chain (mode III fracture) by means of moving forces. The analytical solution was derived together with the expressions that found very good agreement with numerical simulations of the dynamic system.

Firstly, we discussed the quasi-static problem and demonstrated famous lattice trapping phenomenon. Moreover, the expressions for critical force and the energy release rate were obtained. The values for these quantities are shown to be achieved from the dynamic problem when the limit to a stationary crack is taken.

We show the solution of the dynamic problem which was required the application of the Wiener-Hopf technique. The mismatch in material properties of the chains allowed to detect different possibilities. Namely, when the speeds of sound in the chains are different the inter sonic crack speed can appear. Interestingly, there is a gap in the range of the possible crack speeds. To be more precise, no steady-state movement can be detected for the crack speeds grater than the minimum of speeds of sound in separate chains and the less than the speed of sound in the intact part of the double chain.

The solution of the problems allows to evaluate the displacement fields. Their analysis make possible to distinguish the admissible and forbidden steady-state regimes. The forbidden ones are spotted if the fracture condition is checked ahead of the main original crack. For these regimes the special analysis required which lays beyond the scope of the present work. Nevertheless, the predictions on the admissible steady-states can be done carefully with the proposed approach and can be confirmed through the direct numerical integrations of the equations of motion.

The mentioned separation between different steady-state regimes is convenient to illustrate on the energy release rate or force on the crack speed diagrams.
The former allows to estimate the amount of the energy that is carried by elastic waves emanated from a crack tip. It is shown that, as well as in the quasi-static case, the energy release rate is always higher then the energy released when a single spring is broken. Moreover, it allows to predict that the stiffness of the springs connecting to chains effect a significant role in the crack propagation regimes. We show that the slow motion of cracks can be observed when the interfacial springs are relatively compliant. The observations show, that the high crack speeds are always characterised by the high values of energy release rates (low values of energy release rate ratio).

The considerations of force relations shows some intriguing features. The imposed condition on the steady-state made two possible outcomes for the choice of the forces depending on the material properties. If the speeds of sound in two separate chains are the same then the forces should be the same, if their location does not change with time. This condition always happens in the quasi-static formulation of the problem. Distinct values of speeds of sound in the chain provide the values of the forces to be different to avoid the rigid body motion. The intersonic cracks are possible for the same choice of material properties. Moreover, the intersonic crack movement can be provided by means of one force only, that is applied to the chain with a higher value of the speed of sound.

\bibliographystyle{abbrv}
\bibliography{Bibliography}

\end{document}